\documentclass[a4paper,prb,aps,twocolumn,superscriptaddress,floatfix,showkeys]{revtex4-2}
\usepackage[utf8]{inputenc}
\usepackage{graphics}
\usepackage{amsmath}
\usepackage{amsfonts}
\usepackage{hyperref}
\usepackage{cancel}
 \usepackage{multirow}
\usepackage{graphicx}
\usepackage{amssymb}
\usepackage{chemformula}
\usepackage{rotating} 
\usepackage{mathtools} 
\usepackage{siunitx} 
\graphicspath{ {./images/} }

\newcommand{\colvec}[2][.8]{%
  \scalebox{#1}{%
    \renewcommand{\arraystretch}{.8}%
    $\begin{pmatrix}#2\end{pmatrix}$%
  }
}

\newcommand{\nvec}{\boldsymbol{n}}

\begin{document}

\title{Ubiquitous light real-space pairing from long-range hopping and interactions}

\author{G D Adebanjo}
\affiliation{School of Physical Sciences, The Open University, Walton Hall, Milton Keynes, MK7 6AA, UK}

\author{J P Hague}
\affiliation{School of Physical Sciences, The Open University, Walton Hall, Milton Keynes, MK7 6AA, UK}

\author{P E Kornilovitch}
\affiliation{Department of Physics, Oregon State University, Corvallis, OR, 97331, USA}
\date{\today}

\begin{abstract}

We systematically examine how long-range hopping and its synergy with extended interactions leads to light bound pairs. Pair properties are determined for a dilute extended Hubbard model with large on-site repulsion ($U$) and both near- and next-nearest neighbour hopping ($t$ and $t'$) and attraction ($V$ and $V'$), for cubic and tetragonal lattices. The presence of $t'$ and $V'$ promotes light pairs. For tetragonal lattices, $t'<0$ pairs can be lighter than non-interacting particles, and $d$-symmetric pairs form. Close packing transition temperatures, $T^{\ast}$ are estimated for the Bose-Einstein condensation (BEC) of pairs to be $k_{B}T^{\ast}\sim\overline{t} 0.1$, where $\overline{t}$ is the geometric mean of the hoppings on the Cartesian axes. When pairs have $d$-symmetry, the condensate has $d$-wave character. Thus, the presence of both $t'$ and $V'$ leads ubiquitously to small strongly bound pairs with an inverse mass that is linear in hopping, which could lead to high temperature BECs.

\end{abstract}


\maketitle

\section{Introduction}

The properties of real-space pairs are important in the context of scenarios of superconductivity driven by BEC. Real-space pairs have long been known to adequately describe many aspects of the high-Tc phenomenology \cite{Uemura1989,Uemura1991,Alexandrov1994, Alexandrov1999b,Micnas1990}. Recently, real space pairs were detected in iron-based superconductors \cite{Seo2019,Kang2020} and in shot noise in copper oxide tunneling junctions \cite{Zhou2019}. The mass and effective radius of bound pairs define the maximal critical temperature attainable in the system \cite{Ivanov1994,Kornilovitch2015,Zhang2022} making them important quantities to determine. Real-state fermion pairs have also been shown to be stable against further clustering and phase separation \cite{Emery1990,Dagotto1993,Kornilovitch2013,Kornilovitch2014,Chakraborty2014, Kornilovitch2020,Kornilovitch2022}.

In this paper we examine pair properties in dilute extended Hubbard models, concluding that small and light pairs are a ubiquitous consequence of next-nearest neighbor (NNN) interactions and hopping terms in the  Hamiltonian, and discuss their contribution to the high transition temperatures in unconventional superconductors. NNN hopping and NNN interaction are often neglected, but longer range effects may be significant for many materials and lattices. Our goal is to examine the relevance of NNN hopping, $t'$, and interaction, $V'$, to pairs in a range of lattices.

Both the mass and size of real-space pairs contribute to superconducting transition temperatures in the BEC regime. BEC transition temperatures in the dilute limit (where particles rarely scatter) have the form
\begin{align}
	T_{\rm BEC} & = \frac{3.31 \hbar^{2} n^{2/3}_{b} }{k_{B} \overline{m}^{\ast}}.
 \label{eqn:bectc}
\end{align}
In this equation, the geometric mean of the mass, $\overline{m}^{\ast} = (m_{\parallel}^{\ast})^{2/3} (m_{\perp}^{\ast})^{1/3}$ where $m^{\ast}_{\parallel}$ is pair mass in the $xy$ plane, $m^{\ast}_{\perp}$ is pair mass in the $z$ direction (out of the plane), and $n_{b}$ is density of pairs. We use subscript $\parallel$ ($\perp$) to denote properties in the $xy$- ($z$-) directions respoctively. $T_{\rm BEC}$ increases with $n_{b}$ until pairs begin to overlap, at which point $T_{\rm BEC}$ first saturates and then starts to decrease. We estimate that $T_{\rm BEC}$ saturates in the vicinity of {\em close packing}, and define a close packing transition temperature,
\begin{align}
	T^{\ast} & \sim \frac{3.31 \hbar^{2}}{k_{B} \overline{m}^{\ast} \, \Omega^{2/3}_{p}},
 \label{eqn:two}
\end{align}
where $\Omega_{p}$ is pair volume. Substituting for the bare mass for the case of near-neighbor  (NN) hopping only ($m_{0,{\rm NN}}=h^2/(2t a^2)$ in all cases) this becomes:
\begin{equation}
\frac{k_{B} T^{\ast}}{\overline{t}} \sim \frac{6.62}{ \overline{m} \, \Omega_{p}^{\prime 2/3} },
 \label{eqn:twoprime}
\end{equation}
where $\overline{t} = t_{\parallel}^{2/3}t_{\perp}^{1/3}$, $\overline{m} = m_{\parallel}^{2/3} m_{\perp}^{1/3}$ with dimensionless masses $m_{\parallel} = m^{*}_{\parallel} / m_{0,{\rm NN}}$, $m_{\perp} = m^{*}_{\perp} / m_{0,\perp}$ and $\Omega_{p}^{\prime}$ is the pair volume in units of lattice constant $a$ in the $xy$ plane and $b$ in the $z$ direction ($a=b$ for SC and BCC lattices). $m_{0,\perp}=2b^2t$. For tetragonal cases, $t_{\parallel}$ ($t_{\perp}$) is hopping parallel (perpendicular) to the $xy$ plane. For isotropic cases (simple and body centered cubic) $t_{\parallel} = t_{\perp} = t$ and $m_{\parallel} = m_{\perp}$. The precise definition of $\Omega'_{p}$ is subtle, as it needs to be sufficiently large that scattering between pairs is does not significantly affect Eq. \ref{eqn:bectc}. However, some statements can be made: A high $T^{\ast}$ requires compact (small $\Omega_{p}^{\prime}$) and light (small $\overline{m}^{\ast}$) pairs, which in some models leads to a trade off. For example, for large on-site repulsion and NN attraction only, $\Omega_{p}$ decreases with the attractive strength $V$, but at same time the mass increases as $\overline{m}^{\ast} \propto V$ so mass and volume cannot be simultaneously small. It is therefore of interest to examine situations where pair mass and size can be decoupled. In particular, we are interested in systems where the mass remains of the order of bare mass, $\overline{m}^{\ast} \sim m_0$, even in the strong coupling regime, $V \to \infty$ ($m_0$ is defined for different lattices in Tab. \ref{table:one_particle_properties}). We shall refer to such real-space pairs as {\em light}.

\begin{figure*}[t]
    \centering
    \includegraphics[width=175mm]{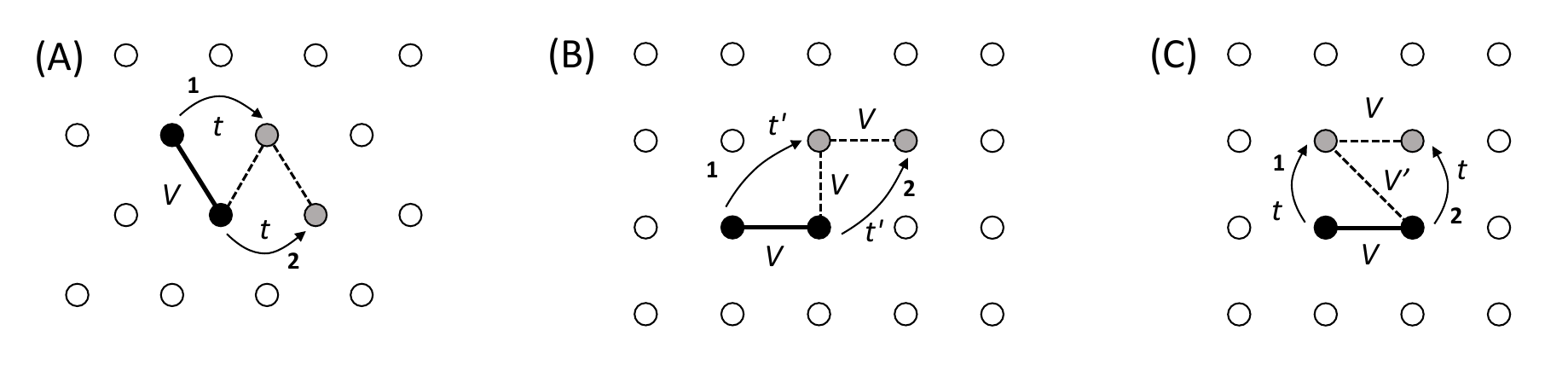}
    \caption{Schematics of different light pair mechanisms. (A) On the triangular lattice, a pair with NN attraction $V$ moves in the first order of NN hopping $t$. (B) On the square lattice, a similar pair with NN attraction requires NNN hopping $t^{\prime}$ to move in the first order. (C) In the resonant case, $V = V^{\prime}$, the pair can move in the first order of NN hopping $t$ even on the square lattice. Numbers `1' and `2' indicate hopping order.}
    \label{fig:schematic}
\end{figure*}

There are several physical mechanisms that can lead to light pairs. In some lattices, the pairs remain light for purely geometric reasons. Consider a triangular lattice with a strong NN attraction $V$, Fig.~\ref{fig:schematic}(A). The pair members can hop in turn without ever breaking the attractive bond. The pair mass is limited to $m^{\ast} < 6 m_0$ even in the $V \to \infty$ limit \cite{hague2007superlighta,hague2008_sing_trip_bip_triangular}. Staggered chains \cite{alexandrov2002b}, staggered square planes \cite{Kornilovitch2023}, and the face-centered-cubic lattice \cite{adebanjofcc2022} are further examples of the geometric mechanism. In other lattices, the geometric mechanism does not work. For example, on a square lattice with only NN attraction and hopping, the pair can only move via an intermediate, higher-energy state. As a result, the pair mass scales as $m^{\ast} \propto m^{2}_{0} V$. The same reasoning applies to simple cubic and body-centered cubic (BCC) lattices with NN $V$ and $t$, and to the attractive Hubbard model on all lattices.

The situation changes qualitatively with the inclusion of NNN hopping, $t^{\prime}$. NNN hopping increases lattice connectivity and the pairs can now move in the {\em first} order in $t^{\prime}$, see Fig.~\ref{fig:schematic}(B). Although in general $t^{\prime} < t$, it is still finite, which means the pair mass no longer scales $\propto V$ but instead saturates at a value defined by $t^{\prime}$. One should note that typically NNN hopping is excluded from a model ``for simplicity'' to make it more tractable. This implicitly assumes that the omission of $t^{\prime}$ does not change basic physics of the system under study. As we have argued, this is not the case for pair mass. $t^{\prime}$ is essential for the physics of real-space pairs and for real-space superconductivity in general. The effects of $t^{\prime}$ are especially prominent in lattices with small NNN-to-NN distance ratios, such as the BCC lattice.     

An even more interesting physics emerges if $t$ and $t^{\prime}$ are of opposite signs. (Typically, $t < 0$ and $t^{\prime} > 0$.) In this case, destructive interference between NN and NNN hopping can lead to a relatively large {\em bare} mass $m_0$. Pairing disrupts the interference by effectively enhancing $t$ or $t^{\prime}$, depending on the details of attraction. With some parameter tuning, one can create a situation where a bound pair is lighter than the bare mass of the constituent particles, $m^{\ast} < 2 m_0$, and even lighter than one bare mass, $m^{\ast} < m_0$. We will refer to such pairs as {\em superlight}.  

Another light pair mechanism is illustrated in Fig. \ref{fig:schematic}(C). In this case, the NNN attraction $V^{\prime}$ is of the same strength as $V$. The pair can move with the NN hopping only because now the intermediate states has the same energy and the mass remains $m^{\ast} \propto m_{0}$. One might think that this mechanism would be rare since it requires fine tuning of parameters. However, in real systems the inter-particle potential is the combination of a short-range repulsion and a longer-range attraction (mediated by phonons or other mechanisms \cite{Callaway1989,Bussmann1989,Zhang1991,Catlow1998,Mihailovich2001,Kabanov2002,Scalapino2012,alexandrov2013SCTHTSC}). Such a potential has a shallow minimum that extends over several lattice constants. Thus, the probability of two potential values being close is not small. 

The focus of this paper is mechanism (B), i.e., the effects of the NNN hopping $t^{\prime}$ on the real-space pair mass. The reason is two-fold. First, this mechanism has not been thoroughly discussed in the literature, unlike mechanisms (A) and (C) \cite{alexandrov2002b,hague2007superlighta,hague2008_sing_trip_bip_triangular,hague2007staggeredladder,adebanjofcc2022,Kornilovitch2023}. Second, mechanism (B) is {\em universal}, i.e., it does not require fine tuning. Any system with a nonzero $t^{\prime}$ will host pairs with $m^{\ast} \propto (t^{\prime})^{-1}$ or lighter. With $t^{\prime}$ taken into account, the lightness of pairs becomes a common and ubiquitous feature of models with intersite attraction, with important implications for unconventional superconductivity discussed earlier. We also investigate interplay between mechanisms (B) and (C) inspired by the high $k_B T_c/\overline{t}$ ratio in BCC fulleride superconductors where $t^{\prime}$ and $V^{\prime}$ are expected to be particularly large.

\section{Model and methodology}
\label{sec:model}
\label{sec:method_uv}

We examine pairs in the dilute limit of an extended Hubbard model which includes NN and longer-range attractive interactions and hopping, with Hamiltonian,
\begin{equation}
\hat{H} = \sum_{{\bf n},{\bf b} \sigma} t_{\bf b} \,
c^{\dagger}_{ {\bf n} + {\bf b} , \sigma } c_{{\bf n}\sigma}  
+ U \sum_{\bf n} \hat{\rho}_{{\bf n} \uparrow} \, \hat{\rho}_{{\bf n} \downarrow}
+ \sum_{{\bf n},{\bf b}} \frac{V_{{\bf b}}}{2} \,
\hat{\rho}_{{\bf n} + {\bf b}} \hat{\rho}_{\bf n}  ,
\label{eqn:hgeneric}    
\end{equation}
where the Hubbard repulsion has been separated out from the other interactions. In the above equation, ${\bf b}$ are the set of vectors to neighbor sites, $c^{\dagger}_{{\bf n}\sigma}$ ($c_{{\bf n}\sigma}$) creates (annihilates) an electron of spin $\sigma$ at site ${\bf n}$, $\hat{\rho}_{{\bf n}\sigma} = c^{\dagger}_{{\bf n}\sigma}c_{{\bf n}\sigma}$, and $\hat{\rho}_{{\bf n}} = \hat{\rho}_{{\bf n}\uparrow} + \hat{\rho}_{{\bf n}\downarrow}$ is the number operator on site ${\bf n}$. One-particle hopping and inter-particle interaction are defined by functions $t_{\bf b}$ (hopping through a displacement ${\bf b}$) and $V_{\bf b}$ (interaction at separation ${\bf b}$), respectively. The factor $\frac{1}{2}$ takes account of double counting of the interaction terms. The models described by Eq.~(\ref{eqn:hgeneric}) are generally known as $UV$ models~\cite{Kornilovitch2023}. 

A two-particle version of Eq.~\ref{eqn:hgeneric} can be found in two cases. The first case is the dilute limit of the Hamilto\-nian. Both electron (mostly negative $t_{\bf b}$'s) and hole (mostly positive $t_{\bf b}$'s) variants have physical meaning. The second, less trivial case occurs close to half filling for the special case $U \gg t_{\bf b}$. Then two holes interact according to Eq. \ref{eqn:hgeneric}, but with renormalized kinetic energy because movement of holes is restricted by strong correlations \cite{Dagotto1994,Leung1995}.\footnote{A dilute Hamiltonian is easily determined, since at very large $U$ and half-filling the wavefunction is a Fock state with 1 particle per site (with constant energy determined by the intersite interactions). When holes are introduced to this Fock state, the difference in the Hamiltonian from the constant background is equivalent to the dilute Hamiltonian. For smaller $U$, such an approximation becomes less good as double occupancy is permitted and increased kinetic energy contributions complicate the half-filled wavefunction.} As a result, we expect that the bands would be flattened and hopping amplitudes of holes (and thus kinetic energy) would be smaller than the $t_{\bf b}$ of the original electron Hamiltonian the hole model is derived from. This increases the relative importance of $V/t$ which could be Coulomb or electron-phonon in origin. A NN interaction can originate from an number of sources. Attractive long-range electron-phonon interactions can lead to instantaneous interactions at large phonon frequency, including an intersite $V$ \cite{hardy2009} (strong phonon mediated intersite attractions have been measured in 1D cuprates \cite{chen2021a,wang_chen_et_al_2021}). In some materials, attractive $V$ were predicted from first-principle quantum chemistry calculations~\cite{Zhang1991,Catlow1998}. Coulomb repulsion can generate a repulsive $V$. The Hubbard $U$ can cause an effective intersite Heisenberg interaction $J$, which while spin dependent could act like $V$ in a dilute system \cite{jozef2017}. There are rich phases in extended Hubbard models, including Mott insulators \cite{hubbard1963electron}, stripe order \cite{kato2000}, $d$-wave superconductivity \cite{PhysRevB.97.184507,hardy2009}, and  XY antiferromagnetism \cite{Laad_1991}. Extended Hubbard models with NN interaction have been applied to superconductors including cuprates \cite{Micnas_1988,PhysRevB.97.184507}.

 We find the exact properties of a pair of spin-$\frac{1}{2}$ fermions in the $UV$ model by solving the following lattice Schr\"odinger equation:
\begin{align}
& \sum_{{\bf b}}t_{{\bf b}}\,\bigg[\Psi({\bf n}_{1}+{{\bf b}},{\bf n}_{2}) + \Psi({\bf n}_{1},{\bf n}_{2}+{\bf b})\bigg]  
\nonumber \\
& + U \delta_{ {\bf n}_{1} , {\bf n}_{2} } \Psi({\bf n}_{1},{\bf n}_{2})
+ \sum_{{\bf b}} {V}_{{\bf b}}\,\delta_{{\bf n}_{1} - {\bf n}_{2},{\bf b}} \Psi({\bf n}_{1},{\bf n}_{2}) 
\nonumber \\
& = E\,\Psi({\bf n}_{1},{\bf n}_{2}) \: ,
\label{eqn:schrodeq}
\end{align}
where $E$ is the total pair energy and $\delta$ is the Kronecker delta function. The pair wavefunction is $\Psi(\nvec_{1},\nvec_{2})$. A complete solution methodology can be found in Ref.~\cite{Kornilovitch2023}. Here we outline key steps and features. (i) In momentum space, Eq.~(\ref{eqn:schrodeq}) becomes an integral equation with a separable kernel. The former is reducible to a system of {\em linear} equations, thus leading to a mathematically exact solution. The size of the linear system is $(1 + n_{\bf b})$ where $n_{\bf b}$ is the number of neighbor vectors with $V_{\bf b} \neq 0$. The system's matrix elements are two-body lattice Green's functions. The solution's complexity grows sharply with the radius of interaction and lattice dimensionality. (ii) By symmetrizing or antisymmetrizing $\Psi({\bf n}_{1},{\bf n}_{2})$, one can focus the solution either on spin-singlet or spin-triplet pair states. Sizes of the respective systems of linear equations are $( 1 + \frac{1}{2} n_{\bf b})$ and $\frac{1}{2} n_{\bf b}$. In most cases, this trick significantly reduces the mathematical complexity of the solution. (iii) The linear system includes pair total momentum ${\bf P}$ as a parameter. Therefore, the exact solution provides pair dispersion $E({\bf P})$ as well as pair mass $m^{\ast}$ defined via a second derivative of $E({\bf P})$. (iv) At the $\Gamma$-point of the pair Brillouin zone, ${\bf P} = 0$, the full linear system can be further decomposed into blocks corresponding to particular orbital symmetries. Sometimes, the mass can also be obtained from the partial blocks. (v) The eigenvector of the linear system defines the pair wave function $\Psi({\bf n}_{1},{\bf n}_{2})$. From there, the pair effective radius and volume can be computed after proper normalization. The final set of linear equations for specific lattices are given in Supplemental Material.  

In addition to the exact solution, we employ an analytical approach to determine pair mass in the strong attraction limit. In this limit, pairs are tightly bound and the basis reduces to a small number of dimer states, which enables a simple derivation of the mass  \cite{hague2007superlighta,hague2008_sing_trip_bip_triangular,hague2007staggeredladder,adebanjofcc2022,Kornilovitch2023}. This mass serves as a useful validation check for the all-coupling solution.

\section{Qualitative considerations}
\label{sec:squaremodel}

In this section, we examine the effects of NNN hopping on real-space pairs in a square $UV$ model with NN attraction. The model serves as a good illustration of the physical ideas explored in this paper and is inspired by the unconventional superconductivity in the underdoped cuprates. Two-dimensional copper-oxygen planes host mobile holes at low densities. The multiorbital model describing these can be mapped to a one-orbital model using a variety of reduction schemes \cite{Zhang1988,Sakakibara2010,Hirayama2018,Jiang2023}. Such procedures typically produce long-range hopping, which we wish to study in this work. An attractive $V$ can originate from several physical mechanisms including mediation by phonons \cite{Alexandrov1994,alexandrov2013SCTHTSC}, spin waves \cite{Scalapino2012}, and the Jahn-Teller effect \cite{Mihailovich2001,Kabanov2002}. An attractive interaction, $V\sim t$ between two holes within a CuO$_2$ plane was inferred from {\em ab initio} quantum chemistry calculations \cite{Zhang1991,Catlow1998}. 

The Hamiltonian we wish to in this section is Eq. \ref{eqn:hgeneric} with $t_{{\bf b}_{1}}=t$, $t_{{\bf b}_{2}}=t'$, $V_{{\bf b}_{1}}=V$, $V_{{\bf b}_{2}}=V'$, 
where ${\bf b}_{1} = \{ \pm {\bf x} ; \pm {\bf y} \}$ and ${\bf b}_{2} = \{ \pm {\bf x} \pm {\bf y} \}$ are the NN and NNN lattice vectors of the square lattice, respectively (${\bf x}$ and ${\bf y}$ are unit vectors on cartesian axes). Note that we defined hopping amplitudes with negative signs. We are mostly interested in the electron channel, $t, t^{\prime} > 0$, although we will also consider $t^{\prime} < 0$ when discussing the superlight effect. The two-particle problem of the square $UV$ model, and of the related $t - J$ model has been considered by several authors \cite{Lin1991,Petukhov1992,Kagan1994,Chernyshev1999,kornilovitch2004}, although the effects of $t^{\prime}$ have not been fully elaborated yet.  

The physical effects discussed below also occur in other lattices with NNN hopping, including the 1D chain and 3D simple cubic lattice. The square $UV$ model serves as a convenient proxy for all hyper-cubic $UV$ models. The non-interacting (one particle) dispersion of Eq.~(\ref{eqn:hgeneric}) is
\begin{align}\label{eq:disp_relation} 
\varepsilon_{\bf k} & = -2t \left( \cos{k_x a} + \cos{k_y a} \right) - 4 t^{\prime} \cos{k_x a} \cos{k_y a} \: ,
\end{align}
where $a$ is the lattice constant, $k_{x}$ and $k_{y}$ are $x$ and $y$ commponents of momentum. The bare effective mass for the square lattice is 
\begin{align} 
m_{0} & = \frac{\hbar^2}{ 2 a^2 ( t + 2 t^{\prime} ) } \: .
\label{eq:sqmass} 
\end{align}

\subsection{Strong-coupling limit and light pairs}
\label{sec:sqstr}

\begin{figure}
    \centering
    \includegraphics[width=70mm]{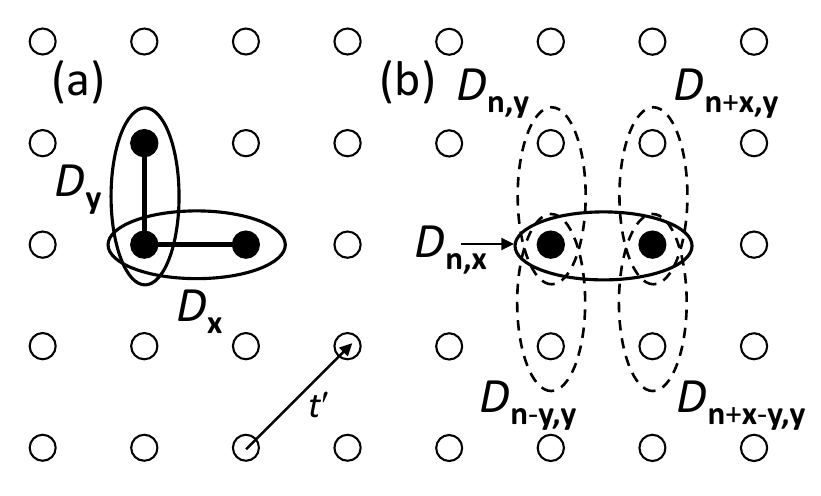}
    \caption{(a) Two types of spin-singlet dimers in the $|V| \gg t, t^{\prime}$ limit of Eq.~(\ref{eqn:hgeneric}). (b) Illustration for the first line in Eq.~(\ref{eq:sqfive}). NNN hopping, $t^{\prime}$, connects a starting dimer of type $\bf x$ with four dimers of type ${\bf y}$. The latter are circled by dashed ovals.}
    \label{fig:dimers}
\end{figure}

We first consider the case of {\em electron}-type NNN hopping with $t^{\prime} > 0$ (in our sign convention).
We begin with considering the strong coupling limit of Eq.~(\ref{eqn:hgeneric}) with NN attraction only, $V^{\prime} = 0$, $V < 0$, and $|V| \gg t, t^{\prime}$. The state basis reduces to tightly bound NN dimers shown in Fig.~\ref{fig:dimers}(a). We consider only spin singlets, hence there are only two dimer types 
\begin{align}
D^{s}_{{\bf n},{\bf b}} & = \frac{1}{\sqrt{2}} \left( 
                \left\vert   \uparrow \right\rangle_{\bf n}
                \left\vert \downarrow \right\rangle_{{\bf n} + {\bf b}}  + 
                \left\vert \downarrow \right\rangle_{\bf n}
                \left\vert   \uparrow \right\rangle_{{\bf n} + {\bf b}} \right) , 
\label{eq:sqdimers}
\end{align}
where ${\bf b} = {\bf x}$ or ${\bf y}$ and $\left|\sigma\right\rangle_{\nvec}=c^{\dagger}_{\nvec\sigma}\left|0\right\rangle$ ($\sigma = \uparrow,\downarrow$). The dimer states have energy $-|V|$ and are connected by NNN hopping. Operating by the $t^{\prime}$ term of $\hat{H}_{\rm sq}$, one obtains, see Fig.~\ref{fig:dimers}(b) 
\begin{align}
    \hat{H}^{\prime}_{\rm sq} D^{s}_{{\bf n},{\bf x}} & = - t^{\prime} \left( D^{s}_{{\bf n},{\bf y}} + D^{s}_{{\bf n}-{\bf y},{\bf y}} + D^{s}_{{\bf n}+{\bf x},{\bf y}} + D^{s}_{{\bf n}+{\bf x}-{\bf y},{\bf y}} \right) 
\nonumber    \\
    \hat{H}^{\prime}_{\rm sq} D^{s}_{{\bf n},{\bf y}} & = - t^{\prime} \left( D^{s}_{{\bf n},{\bf x}} + D^{s}_{{\bf n}+{\bf y},{\bf x}} + D^{s}_{{\bf n}-{\bf x},{\bf x}} + D^{s}_{{\bf n}-{\bf x}+{\bf y},{\bf x}} \right) .
\nonumber    \\
& 
\label{eq:sqfive}    
\end{align}
This leads to a Schr\"odinger equation in momentum space
\begin{align}
& \left[ \begin{array}{cc}
E({\bf P}) + |V| & 
t^{\prime}( 1 + e^{iP_{x}a} )( 1 + e^{-iP_{y}a} ) \\
t^{\prime}( 1 + e^{-iP_{x}a} )( 1 + e^{iP_{y}a} )  & 
E({\bf P}) + |V| 
\end{array} \right] 
\nonumber \\
& \makebox[1.0cm]{} \times \left[ \begin{array}{c}
D^{s}_{{\bf P},{\bf x}} \\
D^{s}_{{\bf P},{\bf y}} 
\end{array} \right] = 0 \: ,
\label{eq:sqsix}    
\end{align}
where ${\bf P}$ is the total pair momentum, with components $P_{x}$ and $P_{y}$. From here, one finds pair dispersion 
\begin{equation} 
E({\bf P}) = - |V| \pm 4 t^{\prime} \cos{\frac{P_{x}a}{2}} \cos{\frac{P_{y}a}{2}} \: ,
\label{eq:sqseven} 
\end{equation}
and pair mass   
\begin{equation} 
m^{\ast} = \frac{\hbar^2}{t^{\prime} a^2} \: . 
\label{eq:sqeight} 
\end{equation}
The pair mass only depends on $t^{\prime}$ and is independent of the binding energy that scales as $|V|$. The pair remains mobile even in the strong coupling limit. This is the {\em light-pair} effect. On the square lattice, it is entirely due to NNN hopping. We also note that since $t^{\prime} > 0$, it follows from Eq.~(\ref{eq:sqsix}) near ${\bf P} = 0$ that $D^{s}_{{\bf P},{\bf x}} = D^{s}_{{\bf P},{\bf y}}$, that is, the pair has $s$-type orbital symmetry.   

When expressed in terms of $m_0$, the pair mass becomes
\begin{equation} 
\left. \frac{m^{\ast}}{m_0} \right\vert_{V \to \infty} 
= \frac{ 2 ( t + 2 t^{\prime} )}{t^{\prime}} \: . 
\label{eq:sqnine} 
\end{equation}
Although $m^{\ast}$ is independent of $V$, the enhancement relative to $m_0$ can still be large if $t^{\prime} \ll t$. The enhancement is minimal when $t^{\prime} \approx t$.

\begin{figure}
    \centering
    \includegraphics[width=70mm]{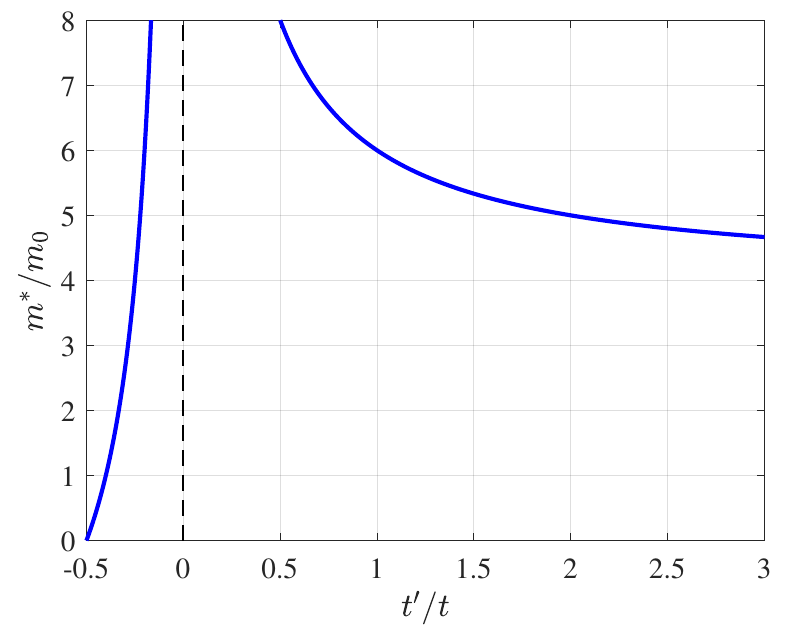}
    \caption{The mass enhancement function, Eq.~(\ref{eq:sqten}), in the $V \gg t, t^{\prime}$ limit. Superlight pairs, $m^{\ast} < m_0$, exist near $t^{\prime}/t = -0.5$. We do not consider the more complex case of $t^{\prime}/t < - 0.5$ where the band minimum is not at ${\bf k} = (0,0)$. }
    \label{fig:superlight}
\end{figure}

\begin{table*}
\begin{tabular}{|l|c|c|c|c|c|c|c|c|c|} \hline
    \multirow{2}{*}{Lattice} & \multirow{2}{*}{$\varepsilon({\bf k})$} & \multirow{2}{*}{$m_{0}$} & \multicolumn{3}{c|}{Distance} & \multicolumn{2}{c|}{$z$} &  \multicolumn{2}{c|}{$E(0)=-W_{0}-W'$}\\ \cline{4-10}
    & & & $d_{\rm NN}$ & $d_{\rm NNN}$ & $\eta$ & $z_{\rm NN}$ & $z_{\rm NNN}$ & $W_{0}$ & $W'$\\\hline
     Square & $-2t(\cos{k_{x}a} + \cos{k_{y}a}) - 4t^{\prime}\cos{k_{x}a}\cos{k_{y}a}$  & $\frac{\hbar^{2}}{2a^{2}(t+2t^{\prime})}$ & $a$ & $\sqrt{2}a$ & $\frac{1}{\sqrt{2}}$ & 4 & 4 & $8t$ &  $ 8t^{\prime}$ \\ \hline
     Tetragonal & $-2t(\cos{k_{x}a} + \cos{k_{y}a}) - 4t^{\prime}\cos{k_{x}a}\cos{k_{y}a} - 2t_{\perp}\cos{k_{z}b}$  & $\frac{\hbar^{2}}{2a^{2}(t+2t^{\prime})}$ & $a$ & $\sqrt{2}a$ & $\frac{1}{\sqrt{2}}$ & 4 & 4 & $8t$ &  $ 8t^{\prime}+4t_{\perp}$ \\ \hline
     \multirow{2}{*}{SC} & $-2t(\cos{k_{x}a} + \cos{k_{y}a} + \cos{k_{z}a})$  & \multirow{2}{*}{$\frac{\hbar^{2}}{2a^{2}(t+4t^{\prime})}$} & \multirow{2}{*}{$a$} & \multirow{2}{*}{$\sqrt{2}a$} & \multirow{2}{*}{$\frac{1}{\sqrt{2}}$} & \multirow{2}{*}{$6$} & \multirow{2}{*}{$12$} & \multirow{2}{*}{$12t$} & \multirow{2}{*}{$ 24t^{\prime}$}\\
     & $- 4t^{\prime}(\cos{k_{x}a}\cos{k_{y}a} + \cos{k_{x}a}\cos{k_{z}a} + \cos{k_{y}a}\cos{k_{z}a} )$ & & & & & & & \\ \hline
     \multirow{2}{*}{BCC} & $-8t \cos \frac{k_{x}a}{2} \cos \frac{k_{y}a}{2} \cos \frac{k_{z}a}{2} $  & \multirow{2}{*}{$\frac{\hbar^{2}}{2a^{2}(t+t^{\prime})}$} & \multirow{2}{*}{$\frac{\sqrt{3}}{2}a$} & \multirow{2}{*}{$a$} & \multirow{2}{*}{$\frac{\sqrt{3}}{2}$} & \multirow{2}{*}{8} & \multirow{2}{*}{6} & \multirow{2}{*}{$16t$} & \multirow{2}{*}{$ 12t^{\prime}$} \\
     & $-2t^{\prime} (\cos k_{x}a + \cos k_{y}a + \cos k_{z}a )$  &  &  &  &  &  & & \\ \hline
\end{tabular}
\caption{Summary of the properties of a particle in various lattices; the lattice constant is $a$, $z$ represents the coordination number, and $\eta=d_{\rm NN}/d_{\rm NNN}$ the ratio between the NN and NNN distances. For tetragonal case, in-plane $m_{0}$ is shown and $b$ is interplane spacing.}\label{table:one_particle_properties}
\end{table*}

\subsection{``Superlight'' bound pairs}
\label{sec:superlight}

We now consider the case of {\em hole}-type NNN hopping, i.e., $t^{\prime} < 0$ with our sign convention. We note that negative $t^{\prime}$ routinely appears in one-orbital models of CuO$_2$ planes, that are derived from multi-orbital models \cite{Sakakibara2010,Hirayama2018,Jiang2023}. Going back to Eq.~(\ref{eq:sqfive}), we now choose the positive sign, which leads to the same ground-state energy $- |V| - 4 |t^{\prime}|$. The pairs mass is still given by Eq.~(\ref{eq:sqeight}) but with $t^{\prime}$ replaced by its absolute value. One important difference concerns the pair symmetry. It follows from Eq.~(\ref{eq:sqsix}) that if $t^{\prime} < 0$, then $D^{s}_{{\bf P},{\bf x}} = - D^{s}_{{\bf P},{\bf y}}$, so the pair has $d$ symmetry. 

Another important change occurs in the one-particle dispersion. It can be seen from Eqs.~(\ref{eq:disp_relation}) and (\ref{eq:sqmass}) that if $t > 0$ and $t^{\prime} < 0$, then destructive interference between NN and NNN hopping flattens the band near ${\bf P} = 0$. As a result, $m_0$ increases and even diverges at $t^{\prime} \to - \frac{1}{2} t$. In the following, we do not consider $t^{\prime} < - \frac{1}{2} t$ because in this parameter region the band bottom shifts from ${\bf P} = 0$ to ${\bf P} = (\pm \pi, 0)$ and mass near the bottom is no longer isotropic. Allowing for negative $t^{\prime}$, Eq.~(\ref{eq:sqnine}) is generalized as follows
\begin{equation} 
\left. \frac{m^{\ast}}{m_0} \right\vert_{V \to \infty} 
= \frac{ 2 ( t + 2 t^{\prime} )}{ | t^{\prime} | } \: ; 
\hspace{0.5cm} t^{\prime} > - \frac{1}{2} t \: . 
\label{eq:sqten} 
\end{equation}
This mass-enhancement function for $V'=0$ is plotted in Fig.~\ref{fig:superlight}. Remarkably, there is a narrow but finite region, $- \frac{1}{2} t < t^{\prime} < - \frac{1}{3} t$, where $m^{\ast} < 2 m_0$, and even a narrower region, $- \frac{1}{2} t < t^{\prime} < - \frac{2}{5} t$, where $m^{\ast} < m_0$, i.e., the tightly bound pair is lighter than one free particle. We call such real-space pairs {\em superlight}. The physics behind the superlight effect is disruption of the destructive interference by interaction: NN attraction elevates the importance of $t^{\prime}$ over $t$. As are result, a bound pair of particles moves more easily than one free particle.

\subsection{Interplay with the resonance mechanism}
\label{sec:sqinter}

In the preceding sections we showed that a nonzero NNN hopping generally leads to light bound pairs and in some special cases to superlight bound pairs. Additionally, resonance in the attractive potential, illustrated in Fig.~\ref{fig:schematic}(c), may also lead to a light mass. In this section we investigate the interplay between the two mechanisms. We consider Eq.~(\ref{eqn:hgeneric}) with an additional condition, $V^{\prime} = V$, and take the limit $V, V^{\prime} \to - \infty$. In addition to the two dimer states shown in Fig.~\ref{fig:dimers}, there will be two more dimer states arranged diagonally. All four dimer types have the same energy and are mixed by $t$ and $t^{\prime}$ hopping. The dimer Schr\"odinger equation is aneigenvalue equation derived in the Supplemental Material. On Brillouin zone diagonals, $P_x = \pm P_y \equiv P$, the full dispersion equation factorizes into the following blocks. (i) Dispersion-less level, $E = 0$. (ii) $d_{x^2-y^2}$ pair with $E = 2 t^{\prime} ( 1 + \cos{P} )$. (iii) $s-d_{xy}$ block with energy defined by  
\begin{equation} 
E^2 + 2 t^{\prime} ( 1 + \cos{P} ) E - 8 t^2 ( 1 + \cos{P} ) = 0 \: . 
\label{eq:sqfifteen} 
\end{equation}
The $s$-symmetric ground state is the lowest root of the above equation
\begin{equation} 
E_{s} = - t^{\prime} ( 1 + \cos{P} ) - 
\sqrt{ t^{\prime 2} ( 1 + \cos{P} )^2 + 8 t^2 ( 1 + \cos{P} ) } \: . 
\label{eq:sqsixteen} 
\end{equation}
Expanding for small $P$, the pair effective mass when $V'=V$ is, 
\begin{equation} 
\frac{m^{\ast}_{s}}{m_{0}} = 
\frac{4 ( t + 2 t^{\prime} ) \sqrt{ t^{\prime 2} + 4 t^2 } }
     { ( t^{\prime 2} + 2 t^2 ) + t^{\prime} \sqrt{ t^{\prime 2} + 4 t^2 } } \: , 
\label{eq:sqseventeen} 
\end{equation}
where $m_{0}$ is given in Eq.~(\ref{eq:sqmass}). $m^{*}/m_{0}$ varies between $4$ and a maximum of $\sim 5.2$, depending on $t'$.

\section{Exact pair properties} \label{sec:results}

In this section, we present pair properties obtained from numerical solution of Eq.~(\ref{eqn:hgeneric}). We study square, quasi-2D tetragonal, 3D simple cubic (SC), and 3D body-centered cubic (BCC) lattices. Study of the tetragonal lattice is inspired by superconductivity in the cuprates while the BCC is inspired by superconductivity in fullerides \cite{martin1993}. In this section, we limit the parameter space to $|V^{\prime}| < |V|$ and $t^{\prime} < t$. The one-particle properties of the model for these systems are summarized in Table~\ref{table:one_particle_properties}. For the tetragonal case, we select $t_{\perp}=0.04$ for hopping between planes. The specific eigenvalue equations are provided in the Supplemental Material.

\begin{figure}
    \centering
    \includegraphics[width=70mm]{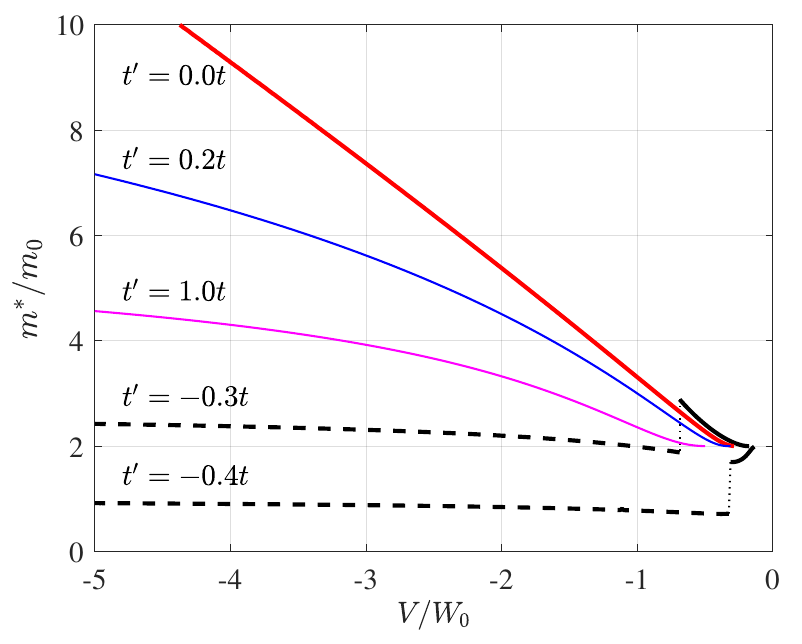}
    \caption{Mass of the ground state bound state in the square $UV$ model. Note that $m_0$ is $t^{\prime}$-dependent, see Eq.~(\ref{eq:sqmass}). Solid lines correspond to $s$-symmetric ground states, with thick and thin lines relating to heavy and light states respectively. Dashed lines correspond to $d$-symmetric superlight states. Notice how the pairs become {\em superlight}, i.e., lighter than the combined mass of the constituent particles, as $t^{\prime} \to -\frac{1}{2} t$. }
    \label{fig:sqmass}
\end{figure}

 Pairs on the square lattice with $V'=0$ and $t'\neq 0$ may be either heavy (with $s$ symmetry), light (also with $s$ symmetry) or superlight (with $d$ symmetry) depending on the parameters. Figure~\ref{fig:sqmass} shows the spin-singlet ground state for a pair of particles on a square lattice. The parameters for the pairing regimes can be categorized as: (i) NN hopping only, $t^{\prime} = 0$. In this ``heavy pair'' regime, the pair only moves in the second order in $t$, hence the mass grows as $m^{\ast} \propto |V|$ to infinity. (ii) Nonzero positive (electron-like) $t^{\prime}$. The mass starts at $m^{\ast} = 2m_0$ at formation. As $|V|$ increases, the mass saturates at a value prescribed by Eq.~(\ref{eq:sqnine}). The pair remains {\em light} and mobile even in the strong coupling limit, $V \to - \infty$, as discussed on Sec.~\ref{sec:sqstr}. The ground state has $s$ orbital symmetry in all regimes. (iii) Finally, there is the {\em superlight} regime, $t^{\prime} < 0$. In this case, there is level crossing. As $|V|$ increases, $s$ pairs are first to form. The mass either decreases or increases, depending of the value of $t^{\prime}$. Upon further increase of $V$, $d$ pairs form and then become the ground state. This effect has been described by Bak and Micnas \cite{Bak1999}. At the crossover, the mass jumps discontinuously. After the crossover, the pairs become superlight (for $t^{\prime}$ close enough to $-\frac{1}{2}t$), i.e., lighter than two separate free particles. In the $V \to - \infty$ limit, the mass approaches Eq.~(\ref{eq:sqten}).         

\begin{figure*}
\includegraphics[width=0.85\textwidth]{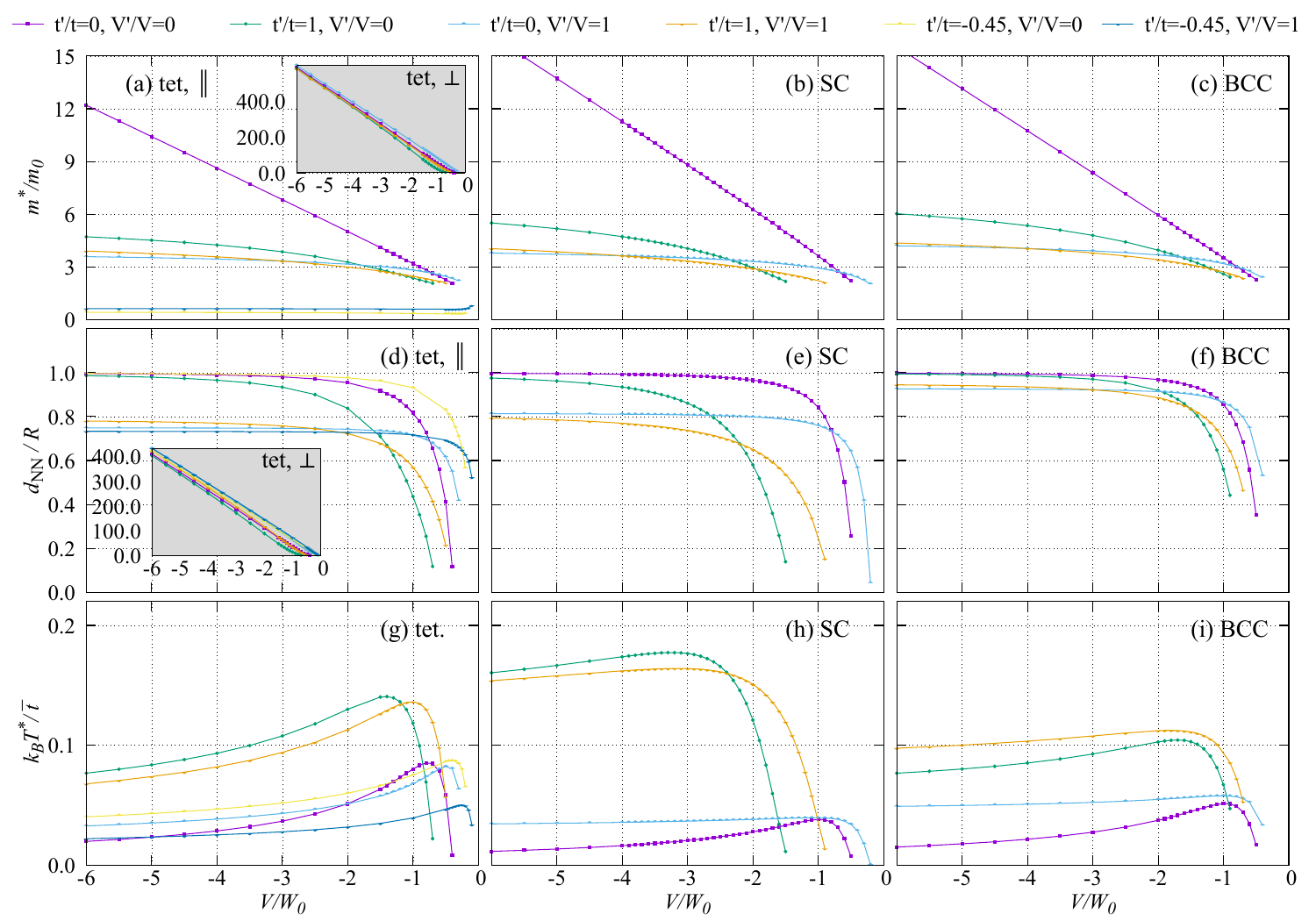}
    \caption{Scaling of mass, inverse radius, and $T^{\ast}$ into the deep potential limit for a subset of the parameter space. The presence of $t'$ or $V'$ leads to light behavior at deep coupling. Such pairs remain lighter in the weak binding limit with small $V$, and higher predicted transition temperatures also persist to weaker binding. For the tetragonal cases, $t_{\perp}/t=0.04$. For strong binding, the radius tends to $R \approx a$ and for the tetragonal case, $R_{\perp}\rightarrow 0$ for the parameters shown. Lines are a guide to the eye.}
    \label{fig:allplots_combined}
\end{figure*}

Moving to 3D lattices, the light pair effect can be demonstrated by examining four extremal cases of the parameter space: (i) $V'/V=0, t'/t=0$, (ii) $V'/V=1, t'/t=0$, (iii) $V'/V=0, t'/t=1$, and (iv) $V'/V=1, t'/t=1$. The additional cases of (v) $V'/V=0, t'/t=-0.45$ and (vi) $V'/V=1, t'/t=-0.45$ are also studied for the tetragonal case. Panels (a)-(c) of Figure~\ref{fig:allplots_combined} 
show how the strong-attraction limit is approached for these cases. When $t'=0$ and $V'=0$, the ``heavy pair'' regime is seen with mass increasing proportionally to $V$. Introduction of $t'$ and $V'$ leads to qualitative changes to the strong coupling mass, which plateaus for strong binding, demonstrating the ``light pair'' regime. The mass reduces with decreasing $V$ until the pairs unbind. Importantly, pairs related to these light asymptotes at strong coupling also remain light at weak binding. Panel (a) shows how the ``superlight'' effect, where pairs are {\em lighter} than the unbound particles is seen for both tetragonal cases with $t'/t=-0.45$. Moreover, in both of those cases, only a small $V$ is required to bind pairs. The $d$-symmetric pairs found in the square lattice persist in the tetragonal lattice for the case of $V'/V=0, t'/t=-0.45$, whereas when $V'/V=1$ pairs have an $s$-symmetry. Out-of-plane masses are large, for the tetragonal lattice, as shown in the inset of Panel (a). The bare mass $m_0$ for a particle for each lattice is given in Table \ref{table:one_particle_properties}. 

The middle row (panels d-f) of Fig.~\ref{fig:allplots_combined} shows the inverse pair radius, calculated as the inverse root mean square (RMS) pair radius. The radius diverges at small $V$ near the pairing threshold. For the smallest $V$ no pairs are formed. In the strong attraction limit ($V \to - \infty$) the radius converges to small values of order $d_{\rm NN}$ if only $V$ is present, or between $d_{\rm NN}$ and $d_{\rm NNN}$ if $V'$ is also present. In the tetragonal case, pairs are confined to planes and the out-of-plane radius is small (much less than a lattice spacing), as shown in the inset of Panel (d). We note that similar masses and radii are found for the chain (not shown here).

The close-packing critical temperature $k_{B}T^{\ast}/\overline{t}$ defined in Eq. \ref{eqn:twoprime} is shown in Panels (g)-(i) of Fig.~\ref{fig:allplots_combined}, and has a maximum value at small to moderate $V$ depending on lattice and model parameters. We select an estimate of $\Omega'_{p}$ that limits scattering by accounting for the exponentially decaying tails of the pair wavefunction ($\Omega'_{p}\sim (\alpha R_{\perp}/b+1)(\alpha R/a+1)^2$, where $R_{\perp}$ ($R_{\parallel}$) is the out-of-plane (in-plane) radius. For cubic lattices, $R_{\perp}=R_{\parallel}=R$ and $b=a$). We note that the location of the maximum in $T^{\ast}$ is insensitive to the value of $\alpha$. For the results shown here we selected $\alpha=5$, to ensure that the exponentially decaying pair tails have very small overlap and scattering is small (essentially eliminating many body effects). A key feature is a maximum in $T^{\ast}$ that occurs between $V \sim -0.1 W_{0}$ and $V\sim -3 W_0$ and is typically broad. The maximum occurs for the following reason: In the low attraction limit, $T^{\ast}$ is dominated by pair radius, with smaller radius leading to higher potential transition temperatures. As attraction increases the pair volume collapses to a minimum before the mass approaches its strong coupling limit, and at this point $T^{\ast}$ is maximal. The maximum occurs at small $V$ for tetragonal cases, since the threshold $V$ for binding is small. Depending on whether the strong attraction limit has light pairs, further increase in $V$ either increases the pair mass, which causes a corresponding gradual decrease of $T^{\ast}$, or the pair mass plateaus, in which case $T^{\ast}$ also plateaus.  For the parameters shown, the overall height of the maximum ranges from $T^{\ast} \sim 0.02t$ to $\sim 0.2t$. Large out-of-plane mass in the tetragonal case limits $T^{\ast}$. These calculations illustrate the potential of real-space superconductivity to reach high critical temperatures if scattering between pairs is small.

One must supplement the estimates of the preceding paragraph with a discussion of {\em phase separation}. Rigorous analysis of pair-pair interaction at the model level \cite{Kornilovitch2022} indicates that real-space pairs can form clusters when $V$ exceeds the pairing threshold by a few $t$s, i.e., a fraction of $W_0$. This implies that the physical part of Fig.~\ref{fig:allplots_combined} is limited to a narrow interval of $V$ close to the binding threshold. In this interval, pair mass is barely above $2m_0$, but the volume consistently decreases and $T^{\ast}$ is increases with $V$ as a result. This suggests a route to higher critical temperatures: increase $V$ to make the pairs more compact but without going over the phase separation threshold. The anisotropy in the tetragonal lattice case may be important in this regard, since deep binding occurs for very small $V$ reducing the risk of crossing the phase separation threshold.

\begin{figure*}
	\centering
	\includegraphics[width=0.95\textwidth]{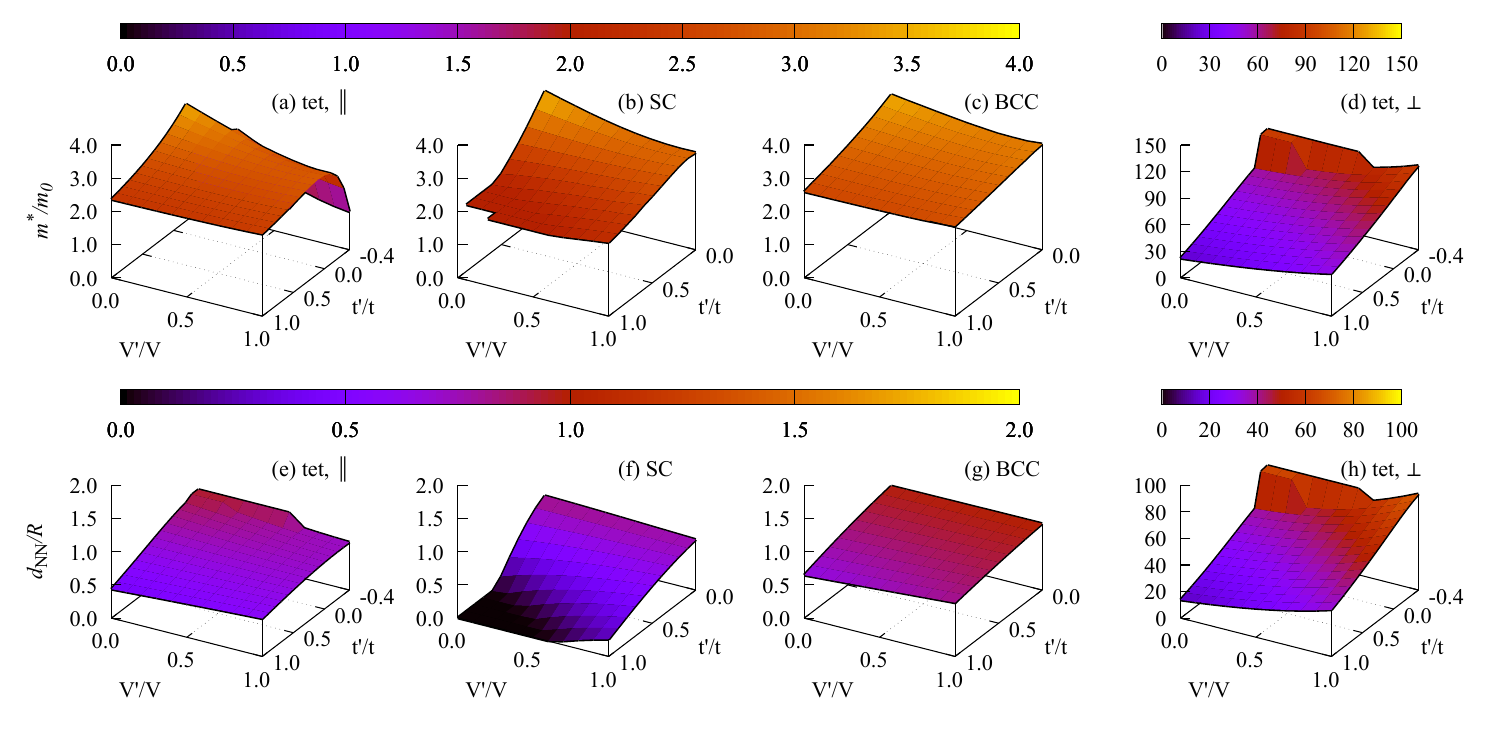}
	\caption{Mass $m^{*}/m_{0}$ and inverse radius $a/R$ for a range of parameters ($V'/V$ and $t'/t$) for fixed $U=2W_{0}$ with $V=-W_{0}$. For the tetragonal case, $t_{\perp}=0.04t$. Panels (a) to (d) show the dependence of the pair mass on $V^{\prime}$ and $t^{\prime}$. Panels (e) to (h) show inverse pair radius. The parameter space is smooth, with the exception of a discontinuity at the boundary between $d$- and $s$-pairing.}
	\label{fig:mass_invrad_tetrag}
\end{figure*}

\begin{figure*}
	\centering
 \includegraphics[width=50mm]{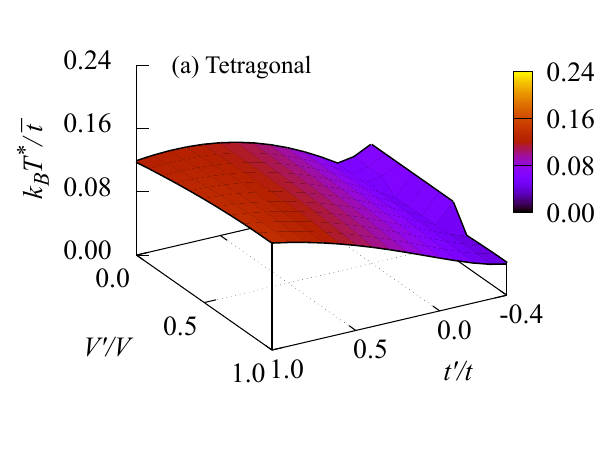}
 \includegraphics[width=50mm]{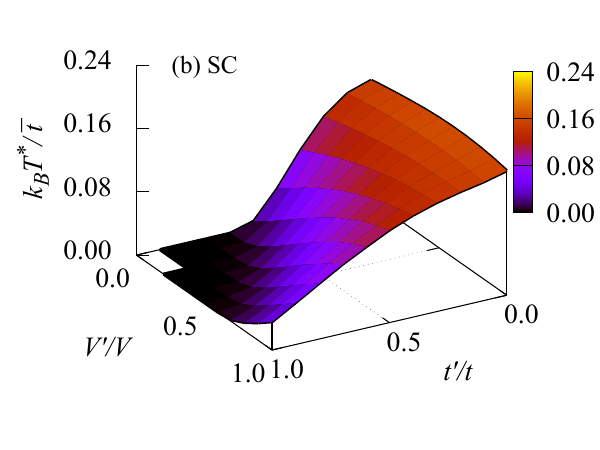}
 \includegraphics[width=50mm]{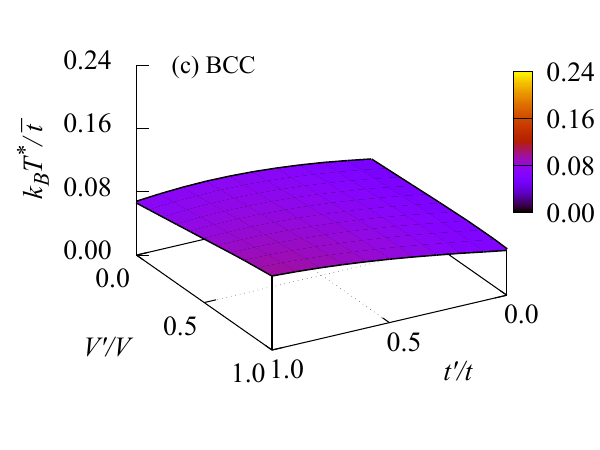}
	\caption{Plot of close packing (maximum) transition temperature from Eqn (\ref{eqn:bectc}) as a function of $V^{\prime}/V$ at varying $t^{\prime}$. For the tetragonal case with $t_{\perp}=0.04$, $d$-symmetry pairs are associated with a small increase in transition temperature.}
	\label{fig:tccp_3D}
\end{figure*}

The presence of $t'$ and $V'$ lead to reductions in the mass, and there is a range of parameters for which pairs are both small and light with masses of just a few single particle masses and radii of a few lattice spacings. Figures \ref{fig:mass_invrad_tetrag} and \ref{fig:tccp_3D} supplement the data of Fig.~\ref{fig:allplots_combined} by showing dependence of mass, radius and $T^{\ast}$ on $V^{\prime}$ and $t^{\prime}$ at a fixed value of $V = -W_{0}$. Panels (a)-(d) of Figure \ref{fig:mass_invrad_tetrag} show the pair mass and Panels (e)-(h) show the inverse radius. We note a jump around $t'\sim 0.4$ in Panels (a),(d),(e) and (h) where the ground state changes from $s$ to $d$ symmetric character in the tetragonal lattice. Otherwise, the plots show smooth interpolation between the four corner cases, demonstrating a lack of special points in the $t^{\prime}-V^{\prime}$ parameter space, and showing that the corner cases are representative of the physical picture above. The inverse radius is typically smallest (radius is largest) for $V'=0$ and $t'/t=1$. In the case of the SC lattice (Panel (f)), the zero inverse radius shows that no bound pair is found in that corner of the plot. In the tetragonal case, pairs are bound in-plane, and in the direction perpendicular to the plane, radius is small and the mass is large, because pairs have to break to move between planes. Moreover, Panel (a) shows the superlight regime of $t'<t$, where pairs are lighter than their component particles. Figure \ref{fig:tccp_3D} shows that close packing transition temperatures vary smoothly, with the exception of the tetragonal lattice where $d$-symmetry pairs are associated with a small and discontinuous increase in transition temperature as levels cross.




\section{Discussion and Conclusions}
\label{sec:four}

In this paper we investigated the factors affecting the masses of real-space bound pairs in models with short-range phenomenological attraction $V$. We showed that inclusion of next-nearest-neighbor (NNN) hopping $t^{\prime}$ limits $m^{\ast}$ to be of the order of bare particle mass $m_0$ on several common lattices. The pairs remain {\em light} even in the strong-coupling limit because they continue moving via resonant configurations without ever breaking the most attractive bond. In cases of tetragonal lattices where $t$ and $t^{\prime}$ are of opposite signs, bound pairs can be {\em superlight}, i.e., lighter than one free particle, $m^{\ast} < m_{0}$. This effect occurs because the negative interference between NN and NNN hopping produces a relatively large $m_0$, whereas coupling between particles disrupts the interference and liberates the pair.

Light and superlight pairs are essential for real-space superconductivity because the BEC temperature scales inversely with $m^{\ast}$. The small and $V$-independent mass offers a recipe for high $T_{\rm BEC}$: systematically increase $V$ to produce more compact pairs. The latter enable higher packing densities and hence higher $T^{\ast}$s. The model calculations of Section~\ref{sec:results} show that the close-packed temperatures can be of the order of tenths of $\overline{t}$, which in physical units amounts to hundreds of degrees. However, this mechanism is limited by phase separation which occurs at intermediate $V$s. Thus, practically achievable $T^{\ast}$ are smaller. Nonetheless, increasing $V$ is a viable path of searching for new high-temperature superconductors. 

In cuprates, $\overline{t}\sim 0.05$ eV (since $t_{\parallel}\sim 0.1$ eV \cite{harrison2024} for holes near half filling and estimating $t_{\perp} \sim 0.04$ eV) so for the cuprates with highest $T_{c}\sim 133K$, $k_{B}T_{c}/\overline{t}\sim 0.15$. Since the pair size in the cuprates is estimated to be 3-10 unit cells from measurements of the upper critical field~\cite{Leggett2006}, the transition temperature is consistent with a close packing picture. Moreover, $d$-symmetric pairs form condensates with a $d$-wave character \cite{bogoliubov}.  Phonon mediated momentum space pairing can also lead to $d$ wave superconductivity: $d$-wave superconducting order parameters can also be found when large $U$ and attractive $V$ with long tails are present \cite{hardy2009} and can also be found in dynamical cluster approximation (DCA) studies of the Holstein model of local electron-phonon coupling \cite{hague2006dwave}. 

We note that the fulleride superconductors are characterized by an unusually high ratio of $k_{B}T_{c}/t$. The BCC fulleride Cs$_{3}$C$_{60}$ has the record transition temperature $T_{c}=38$ K\cite{martin1993}. Models including next-nearest neighbor hopping and interaction may be particularly relevant to BCC lattices, where the ratio of NN to NNN distances is particularly large: $d_{\rm NN}/d_{\rm NNN} \sim 0.866$. This ratio is closer to unity in BCC lattices than in any other Bravais lattices. Thus we expect that $t'/t$ and $V'/V$ is larger than in other Bravais lattices. For fullerides, the approximate hopping value ranges from $t\sim 0.02$ eV to $0.04$ eV (based on the width of $0.61$ eV of the $t_{1u}$ band \cite{gunnarsson2004alkali} and using a range of $t\sim t'$), so $k_{B}T_{c}/t\sim 0.1$ consistent with the close packing estimates here. We advise that further study of many-body effects (such as phase separation) would be required to obtain values for transition temperatures beyond the close-packing estimates and defer this for future work.

\appendix

\section*{Appendix: Supplemental material}

The supplemental material contains specifics related to the solution of the Schr\"odinger equation for specific lattices. It also derives some strong coupling results for the chain, square lattice and BCC lattice.

\begin{widetext}

\section{Solving the Schr\"odinger equation on specific lattices}\label{appendix:uv_derivation}

In this section, we provide information on the solution of the Schr\"odinger equation for pairs on the specific lattices considered in this paper. Refer to Ref.~[\cite{Kornilovitch2023}] for detailed exposition of the procedure. The symmetrized ($+$, spin singlets) and anti-symmetrized ($-$, spin triplets) Schr\"odinger equation in momentum space is given as
\begin{align}\label{total_WF_appendix}
&(E - \varepsilon_{{\bf k}_{1}} - \varepsilon_{{\bf k}_{2}}) \phi_{{\bf k}_{1}{\bf k}_{2}}^{\pm}=\\
&\frac{1}{N}\sideset{}{'} \sum_{{\bf q}{\bf b}}\hat{V}_{{\bf b}}\;\Big\{e^{i({\bf q}-{\bf k}_{1})\,{\bf b}} \pm e^{i({\bf q} - {\bf k}_{2})\,{\bf b}} \Big\}\;\phi_{{\bf q},{\bf k}_{1}+{\bf k}_{2}-{\bf q}}^{\pm} \nonumber .
\end{align}
Here ${\bf b}$ is a lattice vector for which the interaction parameter $\hat{V}_{{\bf b}} \neq 0$. As a result of symmetrization [\cite{Kornilovitch2023}], only one member needs to be retained from any pair $\{ {\bf b} , -{\bf b} \}$. In the problem under study, there are three groups of ${\bf b}$'s: (i) On-site interaction ${\bf b} = {\bf 0}$, $\hat{V}_{{\bf 0}} = U$; (ii) Nearest-neighbor interaction, ${\bf b} = {\bf b}_{1}$, $\hat{V}_{{\bf b}_{1}} = V$, (iii) Next-nearest-neighbor interaction, ${\bf b} = {\bf b}_{2}$, $\hat{V}_{{\bf b}_{2}} = V^{\prime}$. The vectors ${\bf b}_{1}$ and ${\bf b}_{2}$ are listed in Table~\ref{table:apm}. $E$ is the total pair energy and the single particle dispersion $\varepsilon_{\bf k}$ is summarized in Table \ref{table:one_particle_properties}. The prime by the sum sign in Eq.~(\ref{total_WF_appendix}) indicates that the ${\bf b} = {\bf 0}$ term must be taken with an additional factor of $\frac{1}{2}$ in the $(+)$ case and with a factor of $0$ in the $(-)$ case. 

By defining the functions  
\begin{equation}
\Phi_{{\bf b}}^{\pm}({\bf P}) \equiv \frac{1}{N}\sum_{\bf q}e^{i{\bf q}{\bf b}} \; \phi_{{\bf q},{\bf P}-{\bf q}}^{\pm} ,
\end{equation}
where ${\bf P} = {\bf k}_1 + {\bf k}_2$ is the total momentum of the particle pair, the Schr\"odinger equation can be rewritten as an eigenvalue equation
\begin{equation}
\Phi_{{\bf b}}^{\pm}({\bf P})=- \sum_{{\bf b}^{'}}\hat{V}_{{\bf b}}
\mathcal{G}_{{\bf b}{\bf b}^{'}}^{\pm}(E,{\bf P})\;\Phi_{{\bf b}^{'}}^{\pm}({\bf P}) ,
\end{equation}
with matrix elements defined as
\begin{equation}
\mathcal{G}_{{\bf b}{\bf b}^{'}}^{\pm}(E,{\bf P}) = 
\frac{1}{N}\sum_{\bf q}\frac{e^{i{\bf q}({\bf b}-{\bf b}^{'})}\pm e^{i[{\bf q}{\bf b}-({\bf P}-{\bf q}){\bf b}^{'}]}}
{- E + \varepsilon_{{\bf q}} + \varepsilon_{{\bf P}-{\bf q}}} .
\end{equation}
In practice the integrals associated with the matrix elements must be carried out numerically using the Matlab functions \textit{integral}, \textit{integral2} and \textit{integral3}.

Next, we define $\tilde{\Phi}_{\bf b}^{\pm}({\bf P}) = \Phi_{\bf b}^{\pm}({\bf P})e^{-i{\bf P}\cdot{\bf b}/2}$ to simplify the eigenvalue equations. The following subsections list final equations for $E({\bf P})$ for each specific lattice considered in this paper. Equations for spin-singlet and spin-triplets are given separately.

\subsection{1D Chain}

\subsubsection*{Spin singlets}

\begin{gather}\label{eq:1D_sing_arb_P}
\begin{pmatrix}
U\mathcal{G}_0 & V(\mathcal{G}_1 + \mathcal{G}_{\bar{1}})  & V^{\prime}(\mathcal{G}_2 + \mathcal{G}_{\bar{2}})      \\
U\mathcal{G}_1 & V(\mathcal{G}_0 + \mathcal{G}_{2})      & V^{\prime}(\mathcal{G}_{\bar{1}} + \mathcal{G}_3 ) \\
U\mathcal{G}_{2} & V( \mathcal{G}_{1} + \mathcal{G}_{3})      & V^{\prime}(\mathcal{G}_0 + \mathcal{G}_{4})
\end{pmatrix}
\begin{pmatrix}
\tilde{\Phi}^{+}_{0} \\ \tilde{\Phi}^{+}_{1} \\ \tilde{\Phi}^{+}_{2}
\end{pmatrix}
=
\begin{pmatrix}
\tilde{\Phi}^{+}_{0} \\ \tilde{\Phi}^{+}_{1} \\ \tilde{\Phi}^{+}_{2}
\end{pmatrix}
\end{gather}

\subsubsection*{Spin triplets}

\begin{gather}\label{eq:1D_trip_arb_P}
\begin{pmatrix}
 V(\mathcal{G}_0 - \mathcal{G}_{2})      & V^{\prime}(\mathcal{G}_{\bar{1}} - \mathcal{G}_{3}) \\
V(\mathcal{G}_{1} - \mathcal{G}_{3})      & V^{\prime}(\mathcal{G}_0 - \mathcal{G}_{4})
\end{pmatrix}
\begin{pmatrix}
\tilde{\Phi}^{-}_{1} \\ \tilde{\Phi}^{-}_{2}
\end{pmatrix}
=
\begin{pmatrix}
 \tilde{\Phi}^{-}_{1} \\ \tilde{\Phi}^{-}_{2}
\end{pmatrix}
\end{gather}
where 
\begin{align}\label{eq:1D_greens_function_appendix}
    \mathcal{G}_{l}(P) & = \frac{1}{N} \sum_{q}\frac{e^{i l q}}{E - \varepsilon_{\frac{P}{2}+q} - \varepsilon_{\frac{P}{2}-q } } \\
    & = - \int_{-\pi}^{\pi}\frac{dq}{2\pi} \frac{\cos{lq}}{|E| + \varepsilon_{\frac{P}{2} + q} + \varepsilon_{\frac{P}{2} -q}} .
\end{align}
Here and in the following, a bar over subscript $l$ implies a sign change, i.e., $\bar{l} = - l$.  $q$ and $P$ have been scaled so they are dimensionless throughout.

\begin{table*}
\begin{tabular}{|c|p{70mm}|p{85mm}|}
\hline
Lattice & $\{{\bf b}_{1}\}$ & $\{{\bf b}_{2}\}$\\
\hline
Chain     &   $\{ (a) \}$ & $\{ (2a) \}$  \\
\hline
Square & $\{ (a,0), (0,a) \}$ & $\{ (a,a), (a,-a) \}$\\
\hline
SC & $\{ (a,0,0), (0,a,0), (0,0,a) \}$ &  
$\{ (a,a,0), (a,0,a), (0,a,a), (a,-a,0), (a,0,-a), (0,a,-a) \}$ \\
\hline
BCC & $\{ (\frac{a}{2},\frac{a}{2},\frac{a}{2}), (-\frac{a}{2},\frac{a}{2},\frac{a}{2}),(\frac{a}{2},-\frac{a}{2},\frac{a}{2}),(\frac{a}{2},\frac{a}{2},-\frac{a}{2})$ & 
$\{ (a,0,0),(0,a,0), (0,0,a) \}$ \\
\hline
\end{tabular}
\caption{${\bf b}_{1}$ and ${\bf b}_{2}$ for the lattices considered here. $a$ is the lattice constant.}
\label{table:apm}
\end{table*} 
 
\subsection{2D square and 3D tetragonal}

\subsubsection*{Spin singlets}

\begin{gather}\label{eq:2D_sing_arb_P}
\begin{pmatrix}
U\mathcal{G}_{00} & V(\mathcal{G}_{10} + \mathcal{G}_{\bar{1}0}) & V(\mathcal{G}_{01} + \mathcal{G}_{0\bar{1}})  & V^{\prime}(\mathcal{G}_{\bar{1}\bar{1}} + \mathcal{G}_{11}) & V^{\prime}(\mathcal{G}_{\bar{1}1} + \mathcal{G}_{1\bar{1}})     \\
U\mathcal{G}_{10} & V(\mathcal{G}_{00} + \mathcal{G}_{20}) & V(\mathcal{G}_{11} + \mathcal{G}_{1\bar{1}})     & V^{\prime}(\mathcal{G}_{0\bar{1}} + \mathcal{G}_{21} )  & V^{\prime}(\mathcal{G}_{01} + \mathcal{G}_{2\bar{1}} ) \\
U\mathcal{G}_{01} & V(\mathcal{G}_{11} + \mathcal{G}_{\bar{1}1}) & V(\mathcal{G}_{00} + \mathcal{G}_{02})     & V^{\prime}(\mathcal{G}_{\bar{1}0} + \mathcal{G}_{12} )  & V^{\prime}(\mathcal{G}_{10} + \mathcal{G}_{\bar{1}2} ) \\
U\mathcal{G}_{11} & V(\mathcal{G}_{01} + \mathcal{G}_{21}) & V(\mathcal{G}_{10} + \mathcal{G}_{12})     & V^{\prime}(\mathcal{G}_{00} + \mathcal{G}_{22} )  & V^{\prime}(\mathcal{G}_{02} + \mathcal{G}_{20} ) \\
U\mathcal{G}_{1\bar{1}} & V(\mathcal{G}_{0\bar{1}} + \mathcal{G}_{2\bar{1}}) & V(\mathcal{G}_{1\bar{2}} + \mathcal{G}_{10})     & V^{\prime}(\mathcal{G}_{0\bar{2}} + \mathcal{G}_{20} )  & V^{\prime}(\mathcal{G}_{00} + \mathcal{G}_{2\bar{2}} )
\end{pmatrix}
\begin{pmatrix}
\tilde{\Phi}^{+}_{0} \\ \tilde{\Phi}^{+}_{1} \\ \tilde{\Phi}^{+}_{2} \\ \tilde{\Phi}^{+}_{3} \\ \tilde{\Phi}^{+}_{4}
\end{pmatrix}
=
\begin{pmatrix}
\tilde{\Phi}^{+}_{0} \\ \tilde{\Phi}^{+}_{1} \\ \tilde{\Phi}^{+}_{2} \\ \tilde{\Phi}^{+}_{3} \\ \tilde{\Phi}^{+}_{4}
\end{pmatrix}
\end{gather}

\subsubsection*{Spin triplets}

\begin{gather}\label{eq:2D_trip_arb_P}
\begin{pmatrix}
 V(\mathcal{G}_{00} - \mathcal{G}_{20}) & V(\mathcal{G}_{1\bar{1}} - \mathcal{G}_{11})     & V^{\prime}(\mathcal{G}_{0\bar{1}} - \mathcal{G}_{21} )  & V^{\prime}(\mathcal{G}_{01} -\mathcal{G}_{2\bar{1}} ) \\
 V(\mathcal{G}_{\bar{1}1} - \mathcal{G}_{11}) & V(\mathcal{G}_{00} - \mathcal{G}_{02})     & V^{\prime}(\mathcal{G}_{\bar{1}0} - \mathcal{G}_{12} )  & V^{\prime}(\mathcal{G}_{\bar{1}2} - \mathcal{G}_{10}  ) \\
 V(\mathcal{G}_{01} - \mathcal{G}_{21}) & V(\mathcal{G}_{10} - \mathcal{G}_{12})     & V^{\prime}(\mathcal{G}_{00} - \mathcal{G}_{22} )  & V^{\prime}(\mathcal{G}_{02} - \mathcal{G}_{20} ) \\
V(\mathcal{G}_{0\bar{1}} - \mathcal{G}_{2\bar{1}}) & V(\mathcal{G}_{1\bar{2}} - \mathcal{G}_{10})     & V^{\prime}(\mathcal{G}_{0\bar{2}} - \mathcal{G}_{20} )  & V^{\prime}(\mathcal{G}_{00} - \mathcal{G}_{2\bar{2}} )
\end{pmatrix}
\begin{pmatrix}
\tilde{\Phi}^{-}_{1} \\ \tilde{\Phi}^{-}_{2} \\ \tilde{\Phi}^{-}_{3} \\ \tilde{\Phi}^{-}_{4}
\end{pmatrix}
=
\begin{pmatrix}
 \tilde{\Phi}^{-}_{1} \\ \tilde{\Phi}^{-}_{2} \\ \tilde{\Phi}^{-}_{3} \\ \tilde{\Phi}^{-}_{4}
\end{pmatrix}
\end{gather}
where 
\begin{equation}\label{eq:2D_greens_function_appendix}
    \mathcal{G}_{lm}({\bf P})=\frac{1}{N} \sum_{\bf q}\frac{e^{i(lq_{x} + mq_{y})}}{E-\varepsilon_{\frac{\bf P}{2}+{\bf q}}-\varepsilon_{\frac{\bf P}{2}-{\bf q}}} = - \int_{-\pi}^{\pi}\int_{-\pi}^{\pi}\frac{dq_{x} dq_{y}}{(2\pi)^{2}}\frac{\cos ( lq_{x} + mq_{y} )}{|E|+\varepsilon_{\frac{\bf P}{2} + {\bf q}} + \varepsilon_{\frac{\bf P}{2} - {\bf q}}} \: .
\end{equation}
and the tetragonal case is similar, except that the integral is over three dimensions and the dispersion, $\varepsilon$ differs from the square case to take account of the hopping between planes (see Table \ref{table:one_particle_properties}):
\begin{equation}\label{eq:3D_greens_function_appendix_tetragonal}
    \mathcal{G}_{lm}({\bf P})= - \int_{-\pi}^{\pi}\int_{-\pi}^{\pi}\int_{-\pi}^{\pi}\frac{dq_{x} dq_{y} dq_{z}}{(2\pi)^{3}}\frac{\cos ( lq_{x} + mq_{y} )}{|E|+\varepsilon_{\frac{\bf P}{2} + {\bf q}} + \varepsilon_{\frac{\bf P}{2} - {\bf q}}} \: .
\end{equation}
Again, $q$ and $P$ are dimensionless here.

\subsection{3D Simple Cubic (SC)}

Due to the size of the matrices in the simple cubic case, we have not included  the column matrices, $\tilde{\Phi}^{\pm}_{n} $. The process of finding pair energies consists of varying $E$ until one or more eigenvalues of the following matrix equals 1.

\subsubsection*{Spin singlets}

\begin{equation}\label{eq:SC_sing_arb_P}
 \colvec[.72]{
U\mathcal{G}_{000} & V(\mathcal{G}_{\bar{1}00} + \mathcal{G}_{100} ) & V(\mathcal{G}_{0\bar{1}0} + \mathcal{G}_{010})  & V(\mathcal{G}_{00\bar{1}} + \mathcal{G}_{001}) & V^{\prime}(\mathcal{G}_{\bar{1}\bar{1}0} + \mathcal{G}_{110}) & V^{\prime}(\mathcal{G}_{\bar{1}0\bar{1}} + \mathcal{G}_{101}) & V^{\prime}(\mathcal{G}_{0\bar{1}\bar{1}} + \mathcal{G}_{011}) & V^{\prime}(\mathcal{G}_{\bar{1}10} + \mathcal{G}_{1\bar{1}0}) & V^{\prime}(\mathcal{G}_{\bar{1}01} + \mathcal{G}_{10\bar{1}}) & V^{\prime}(\mathcal{G}_{0\bar{1}1} + \mathcal{G}_{01\bar{1}})  
\\
U\mathcal{G}_{100} & V(\mathcal{G}_{000} + \mathcal{G}_{200}) & V(\mathcal{G}_{1\bar{1}0} + \mathcal{G}_{110}) & V(\mathcal{G}_{10\bar{1}} + \mathcal{G}_{101}) & V^{\prime}(\mathcal{G}_{0\bar{1}0} + \mathcal{G}_{210} )  & V^{\prime}(\mathcal{G}_{00\bar{1}} + \mathcal{G}_{201} ) & V^{\prime}(\mathcal{G}_{1\bar{1}\bar{1}} + \mathcal{G}_{111} ) & V^{\prime}(\mathcal{G}_{010} + \mathcal{G}_{2\bar{}01} ) & V^{\prime}(\mathcal{G}_{001} + \mathcal{G}_{20\bar{1}} ) & V^{\prime}(\mathcal{G}_{1\bar{1}1} + \mathcal{G}_{11\bar{1}} )
\\
U\mathcal{G}_{010} & V(\mathcal{G}_{\bar{1}10} + \mathcal{G}_{110}) & V(\mathcal{G}_{000} + \mathcal{G}_{020})  & V(\mathcal{G}_{01\bar{1}} + \mathcal{G}_{011})    & V^{\prime}(\mathcal{G}_{\bar{1}00} + \mathcal{G}_{120} )  & V^{\prime}(\mathcal{G}_{\bar{1}1\bar{1}} + \mathcal{G}_{111} ) & V^{\prime}(\mathcal{G}_{00\bar{1}} + \mathcal{G}_{021} ) & V^{\prime}(\mathcal{G}_{\bar{1}20} + \mathcal{G}_{100} ) & V^{\prime}(\mathcal{G}_{\bar{1}11} + \mathcal{G}_{11\bar{1}} ) & V^{\prime}(\mathcal{G}_{001} + \mathcal{G}_{02\bar{1}} )
\\
U\mathcal{G}_{001} & V(\mathcal{G}_{\bar{1}01} + \mathcal{G}_{101}) & V(\mathcal{G}_{0\bar{1}1} + \mathcal{G}_{011}) & V(\mathcal{G}_{000} + \mathcal{G}_{002})      & V^{\prime}(\mathcal{G}_{\bar{1}\bar{1}1} + \mathcal{G}_{111} )  & V^{\prime}(\mathcal{G}_{\bar{1}00} + \mathcal{G}_{102} ) & V^{\prime}(\mathcal{G}_{0\bar{1}0} + \mathcal{G}_{012} ) & V^{\prime}(\mathcal{G}_{\bar{1}11} + \mathcal{G}_{1\bar{1}1} ) & V^{\prime}(\mathcal{G}_{\bar{1}02} + \mathcal{G}_{100} ) & V^{\prime}(\mathcal{G}_{0\bar{1}2} + \mathcal{G}_{010} )
\\
U\mathcal{G}_{110} & V(\mathcal{G}_{010} + \mathcal{G}_{210}) & V(\mathcal{G}_{100} + \mathcal{G}_{120}) & V(\mathcal{G}_{11\bar{1}} + \mathcal{G}_{111})     & V^{\prime}(\mathcal{G}_{000} + \mathcal{G}_{220} )  & V^{\prime}(\mathcal{G}_{01\bar{1}} + \mathcal{G}_{211} ) & V^{\prime}(\mathcal{G}_{10\bar{1}} + \mathcal{G}_{121} ) & V^{\prime}(\mathcal{G}_{020} + \mathcal{G}_{200} ) & V^{\prime}(\mathcal{G}_{011} + \mathcal{G}_{21\bar{1}} ) & V^{\prime}(\mathcal{G}_{101} + \mathcal{G}_{12\bar{1}} )
\\
U\mathcal{G}_{101} & V(\mathcal{G}_{001} + \mathcal{G}_{201}) & V(\mathcal{G}_{1\bar{1}1} + \mathcal{G}_{111}) & V(\mathcal{G}_{100} + \mathcal{G}_{102})    & V^{\prime}(\mathcal{G}_{0\bar{1}1} + \mathcal{G}_{211} )  & V^{\prime}(\mathcal{G}_{000} + \mathcal{G}_{202} ) & V^{\prime}(\mathcal{G}_{1\bar{1}0} + \mathcal{G}_{112} ) & V^{\prime}(\mathcal{G}_{011} + \mathcal{G}_{2\bar{1}1} ) & V^{\prime}(\mathcal{G}_{002} + \mathcal{G}_{200} ) & V^{\prime}(\mathcal{G}_{1\bar{1}2} + \mathcal{G}_{110} )
\\
U\mathcal{G}_{011} & V(\mathcal{G}_{\bar{1}11} + \mathcal{G}_{111}) & V(\mathcal{G}_{001} + \mathcal{G}_{021}) & V(\mathcal{G}_{010} + \mathcal{G}_{012})    & V^{\prime}(\mathcal{G}_{\bar{1}01} + \mathcal{G}_{121} )  &  V^{\prime}(\mathcal{G}_{\bar{1}10} + \mathcal{G}_{112} ) & V^{\prime}(\mathcal{G}_{000} + \mathcal{G}_{022} ) & V^{\prime}(\mathcal{G}_{\bar{1}21} + \mathcal{G}_{101} ) & V^{\prime}(\mathcal{G}_{\bar{1}12} + \mathcal{G}_{110} ) & V^{\prime}(\mathcal{G}_{002} + \mathcal{G}_{020} )
\\
U\mathcal{G}_{1\bar{1}0} & V(\mathcal{G}_{0\bar{1}0} + \mathcal{G}_{2\bar{1}0}) & V(\mathcal{G}_{1\bar{2}0} + \mathcal{G}_{100}) & V(\mathcal{G}_{1\bar{1}\bar{1}} + \mathcal{G}_{1\bar{1}1})    & V^{\prime}(\mathcal{G}_{0\bar{2}0} + \mathcal{G}_{200} )  &  V^{\prime}(\mathcal{G}_{0\bar{1}\bar{1}} + \mathcal{G}_{2\bar{1}1} ) & V^{\prime}(\mathcal{G}_{1\bar{2}\bar{1}} + \mathcal{G}_{101} ) & V^{\prime}(\mathcal{G}_{000} + \mathcal{G}_{2\bar{2}0} ) & V^{\prime}(\mathcal{G}_{0\bar{1}1} + \mathcal{G}_{2\bar{1}\bar{1}} ) & V^{\prime}(\mathcal{G}_{1\bar{2}1} + \mathcal{G}_{10\bar{1}} )
\\
U\mathcal{G}_{10\bar{1}} & V(\mathcal{G}_{00\bar{1}} + \mathcal{G}_{20\bar{1}}) & V(\mathcal{G}_{1\bar{1}\bar{1}} + \mathcal{G}_{11\bar{1}}) & V(\mathcal{G}_{10\bar{2}} + \mathcal{G}_{100})    & V^{\prime}(\mathcal{G}_{0\bar{1}\bar{1}} + \mathcal{G}_{21\bar{1}} )  &  V^{\prime}(\mathcal{G}_{00\bar{2}} + \mathcal{G}_{200} ) & V^{\prime}(\mathcal{G}_{1\bar{1}\bar{2}} + \mathcal{G}_{110} ) &  V^{\prime}(\mathcal{G}_{01\bar{1}} + \mathcal{G}_{2\bar{1}\bar{1}} ) & V^{\prime}(\mathcal{G}_{000} + \mathcal{G}_{20\bar{2}} ) & V^{\prime}(\mathcal{G}_{1\bar{1}0} + \mathcal{G}_{11\bar{2}} )
\\
U\mathcal{G}_{01\bar{1}} & V(\mathcal{G}_{\bar{1}1\bar{1}} + \mathcal{G}_{11\bar{1}}) & V(\mathcal{G}_{00\bar{1}} + \mathcal{G}_{02\bar{1}}) & V(\mathcal{G}_{01\bar{2}} + \mathcal{G}_{010})    & V^{\prime}(\mathcal{G}_{\bar{1}0\bar{1}} + \mathcal{G}_{12\bar{1}} )  &  V^{\prime}(\mathcal{G}_{\bar{1}1\bar{2}} + \mathcal{G}_{110} ) & V^{\prime}(\mathcal{G}_{00\bar{2}} + \mathcal{G}_{020} ) &  V^{\prime}(\mathcal{G}_{\bar{1}2\bar{1}} + \mathcal{G}_{10\bar{1}} ) & V^{\prime}(\mathcal{G}_{\bar{1}10} + \mathcal{G}_{11\bar{2}} ) & V^{\prime}(\mathcal{G}_{000} + \mathcal{G}_{02\bar{2}} ) 
}
\end{equation}

\subsubsection*{Spin triplets}

\begin{equation}\label{eq:SC_trip_arb_P}
 \colvec[.74]{
 V(\mathcal{G}_{000} - \mathcal{G}_{200}) & V(\mathcal{G}_{1\bar{1}0} - \mathcal{G}_{110}) & V(\mathcal{G}_{10\bar{1}} - \mathcal{G}_{101}) & V^{\prime}(\mathcal{G}_{0\bar{1}0} - \mathcal{G}_{210} )  & V^{\prime}(\mathcal{G}_{00\bar{1}} - \mathcal{G}_{201} ) & V^{\prime}(\mathcal{G}_{1\bar{1}\bar{1}} - \mathcal{G}_{111} ) & V^{\prime}(\mathcal{G}_{010} - \mathcal{G}_{2\bar{}01} ) & V^{\prime}(\mathcal{G}_{001} - \mathcal{G}_{20\bar{1}} ) & V^{\prime}(\mathcal{G}_{1\bar{1}1} - \mathcal{G}_{11\bar{1}} )
\\
 V(\mathcal{G}_{\bar{1}10} - \mathcal{G}_{110}) & V(\mathcal{G}_{000} - \mathcal{G}_{020})  & V(\mathcal{G}_{01\bar{1}} - \mathcal{G}_{011})    & V^{\prime}(\mathcal{G}_{\bar{1}00} - \mathcal{G}_{120} )  & V^{\prime}(\mathcal{G}_{\bar{1}1\bar{1}} - \mathcal{G}_{111} ) & V^{\prime}(\mathcal{G}_{00\bar{1}} - \mathcal{G}_{021} ) & V^{\prime}(\mathcal{G}_{\bar{1}20} - \mathcal{G}_{100} ) & V^{\prime}(\mathcal{G}_{\bar{1}11} - \mathcal{G}_{11\bar{1}} ) & V^{\prime}(\mathcal{G}_{001} - \mathcal{G}_{02\bar{1}} )
\\
 V(\mathcal{G}_{\bar{1}01} - \mathcal{G}_{101}) & V(\mathcal{G}_{0\bar{1}1} - \mathcal{G}_{011}) & V(\mathcal{G}_{000} - \mathcal{G}_{002})      & V^{\prime}(\mathcal{G}_{\bar{1}\bar{1}1} - \mathcal{G}_{111} )  & V^{\prime}(\mathcal{G}_{\bar{1}00} - \mathcal{G}_{102} ) & V^{\prime}(\mathcal{G}_{0\bar{1}0} - \mathcal{G}_{012} ) & V^{\prime}(\mathcal{G}_{\bar{1}11} - \mathcal{G}_{1\bar{1}1} ) & V^{\prime}(\mathcal{G}_{\bar{1}02} - \mathcal{G}_{100} ) & V^{\prime}(\mathcal{G}_{0\bar{1}2} - \mathcal{G}_{010} )
\\
 V(\mathcal{G}_{010} - \mathcal{G}_{210}) & V(\mathcal{G}_{100} - \mathcal{G}_{120}) & V(\mathcal{G}_{11\bar{1}} - \mathcal{G}_{111})     & V^{\prime}(\mathcal{G}_{000} - \mathcal{G}_{220} )  & V^{\prime}(\mathcal{G}_{01\bar{1}} - \mathcal{G}_{211} ) & V^{\prime}(\mathcal{G}_{10\bar{1}} - \mathcal{G}_{121} ) & V^{\prime}(\mathcal{G}_{020} - \mathcal{G}_{200} ) & V^{\prime}(\mathcal{G}_{011} - \mathcal{G}_{21\bar{1}} ) & V^{\prime}(\mathcal{G}_{101} - \mathcal{G}_{12\bar{1}} )
\\
 V(\mathcal{G}_{001} - \mathcal{G}_{201}) & V(\mathcal{G}_{1\bar{1}1} - \mathcal{G}_{111}) & V(\mathcal{G}_{100} - \mathcal{G}_{102})    & V^{\prime}(\mathcal{G}_{0\bar{1}1} - \mathcal{G}_{211} )  & V^{\prime}(\mathcal{G}_{000} - \mathcal{G}_{202} ) & V^{\prime}(\mathcal{G}_{1\bar{1}0} - \mathcal{G}_{112} ) & V^{\prime}(\mathcal{G}_{011} - \mathcal{G}_{2\bar{1}1} ) & V^{\prime}(\mathcal{G}_{002} - \mathcal{G}_{200} ) & V^{\prime}(\mathcal{G}_{1\bar{1}2} - \mathcal{G}_{110} )
\\
 V(\mathcal{G}_{\bar{1}11} - \mathcal{G}_{111}) & V(\mathcal{G}_{001} - \mathcal{G}_{021}) & V(\mathcal{G}_{010} - \mathcal{G}_{012})    & V^{\prime}(\mathcal{G}_{\bar{1}01} - \mathcal{G}_{121} )  &  V^{\prime}(\mathcal{G}_{\bar{1}10} - \mathcal{G}_{112} ) & V^{\prime}(\mathcal{G}_{000} - \mathcal{G}_{022} ) & V^{\prime}(\mathcal{G}_{\bar{1}21} - \mathcal{G}_{101} ) & V^{\prime}(\mathcal{G}_{\bar{1}12} - \mathcal{G}_{110} ) & V^{\prime}(\mathcal{G}_{002} - \mathcal{G}_{020} )
\\
 V(\mathcal{G}_{0\bar{1}0} - \mathcal{G}_{2\bar{1}0}) & V(\mathcal{G}_{1\bar{2}0} - \mathcal{G}_{100}) & V(\mathcal{G}_{1\bar{1}\bar{1}} - \mathcal{G}_{1\bar{1}1})    & V^{\prime}(\mathcal{G}_{0\bar{2}0} - \mathcal{G}_{200} )  &  V^{\prime}(\mathcal{G}_{0\bar{1}\bar{1}} - \mathcal{G}_{2\bar{1}1} ) & V^{\prime}(\mathcal{G}_{1\bar{2}\bar{1}} - \mathcal{G}_{101} ) & V^{\prime}(\mathcal{G}_{000} - \mathcal{G}_{2\bar{2}0} ) & V^{\prime}(\mathcal{G}_{0\bar{1}1} - \mathcal{G}_{2\bar{1}\bar{1}} ) & V^{\prime}(\mathcal{G}_{1\bar{2}1} - \mathcal{G}_{10\bar{1}} )
\\
 V(\mathcal{G}_{00\bar{1}} - \mathcal{G}_{20\bar{1}}) & V(\mathcal{G}_{1\bar{1}\bar{1}} - \mathcal{G}_{11\bar{1}}) & V(\mathcal{G}_{10\bar{2}} - \mathcal{G}_{100})    & V^{\prime}(\mathcal{G}_{0\bar{1}\bar{1}} - \mathcal{G}_{21\bar{1}} )  &  V^{\prime}(\mathcal{G}_{00\bar{2}} - \mathcal{G}_{200} ) & V^{\prime}(\mathcal{G}_{1\bar{1}\bar{2}} - \mathcal{G}_{110} ) &  V^{\prime}(\mathcal{G}_{01\bar{1}} - \mathcal{G}_{2\bar{1}\bar{1}} ) & V^{\prime}(\mathcal{G}_{000} - \mathcal{G}_{20\bar{2}} ) & V^{\prime}(\mathcal{G}_{1\bar{1}0} - \mathcal{G}_{11\bar{2}} )
\\
 V(\mathcal{G}_{\bar{1}1\bar{1}} - \mathcal{G}_{11\bar{1}}) & V(\mathcal{G}_{00\bar{1}} - \mathcal{G}_{02\bar{1}}) & V(\mathcal{G}_{01\bar{2}} - \mathcal{G}_{010})    & V^{\prime}(\mathcal{G}_{\bar{1}0\bar{1}} - \mathcal{G}_{12\bar{1}} )  &  V^{\prime}(\mathcal{G}_{\bar{1}1\bar{2}} - \mathcal{G}_{110} ) & V^{\prime}(\mathcal{G}_{00\bar{2}} - \mathcal{G}_{020} ) &  V^{\prime}(\mathcal{G}_{\bar{1}2\bar{1}} - \mathcal{G}_{10\bar{1}} ) & V^{\prime}(\mathcal{G}_{\bar{1}10} - \mathcal{G}_{11\bar{2}} ) & V^{\prime}(\mathcal{G}_{000} - \mathcal{G}_{02\bar{2}} ) 
}
\end{equation}
where 
\begin{equation}\label{eq:SC_greens_function_appendix}
    \mathcal{G}_{lmn}({\bf P}) = \frac{1}{N} \sum_{\bf q} \frac{ e^{i(lq_{x} + mq_{y} + nq_{z})} }
    {E - \varepsilon_{\frac{\bf P}{2}+{\bf q}} - 
         \varepsilon_{\frac{\bf P}{2}-{\bf q}}} 
= - \int_{-\pi}^{\pi} \int_{-\pi}^{\pi} \int_{-\pi}^{\pi} \frac{dq_{x}dq_{y}}{(2\pi)^{3}}\frac{\cos ( lq_{x} + mq_{y} + nq_{z} )}{|E| + \varepsilon_{\frac{\bf P}{2}+{\bf q}} + \varepsilon_{\frac{\bf P}{2}-{\bf q}}} \: .
\end{equation}

\subsection{3D Body-Centered Cubic (BCC)}

\subsubsection*{Spin singlets}

\begin{equation}\label{appendix_singlet_sel_cons_eqn}
    \colvec[0.8]{
    U\mathcal{G}_{000} & V(\mathcal{G}_{111}+\mathcal{G}_{\bar{1}\bar{1}\bar{1}}) & V(\mathcal{G}_{1\bar{1}\bar{1}}+\mathcal{G}_{\bar{1}11} ) & V(\mathcal{G}_{1\bar{1}1}+\mathcal{G}_{\bar{1}1\bar{1}}) & V(\mathcal{G}_{11\bar{1}}+\mathcal{G}_{\bar{1}\bar{1}1}) & V^{\prime} (\mathcal{G}_{\bar{2}00}+\mathcal{G}_{200}) & V^{\prime}(\mathcal{G}_{0\bar{2}0}+\mathcal{G}_{020}) & V^{\prime}(\mathcal{G}_{00\bar{2}}+\mathcal{G}_{002}) \\
    \\
	U\mathcal{G}_{111} & V(\mathcal{G}_{000}+\mathcal{G}_{222}) & V(\mathcal{G}_{200} + \mathcal{G}_{022}) & V(\mathcal{G}_{020} + \mathcal{G}_{202}) & V(\mathcal{G}_{002} + \mathcal{G}_{220}) & V^{\prime} (\mathcal{G}_{\bar{1}11}+\mathcal{G}_{311}) & V^{\prime}(\mathcal{G}_{1\bar{1}1}+\mathcal{G}_{131}) & V^{\prime}(\mathcal{G}_{11\bar{1}}+\mathcal{G}_{113}) \\
	\\
	U\mathcal{G}_{\bar{1}11} & V(\mathcal{G}_{\bar{2}00} + \mathcal{G}_{022}) & V(\mathcal{G}_{000} + \mathcal{G}_{\bar{2}22}) & V(\mathcal{G}_{\bar{2}20}+\mathcal{G}_{002}) & V(\mathcal{G}_{\bar{2}02} + \mathcal{G}_{020}) & V^{\prime} (\mathcal{G}_{\bar{3}11}+\mathcal{G}_{111}) & V^{\prime}(\mathcal{G}_{\bar{1}\bar{1}1}+\mathcal{G}_{\bar{1}31}) & V^{\prime}(\mathcal{G}_{\bar{1}1\bar{1}}+\mathcal{G}_{\bar{1}13}) \\
	\\
	U\mathcal{G}_{1\bar{1}1} & V(\mathcal{G}_{0\bar{2}0} + \mathcal{G}_{202}) & V(\mathcal{G}_{2\bar{2}0} + \mathcal{G}_{002}) & V(\mathcal{G}_{000} + \mathcal{G}_{2\bar{2}2}) & V(\mathcal{G}_{0\bar{2}2} + \mathcal{G}_{200}) & V^{\prime} (\mathcal{G}_{\bar{1}\bar{1}1}+\mathcal{G}_{3\bar{1}1}) & V^{\prime}(\mathcal{G}_{1\bar{3}1}+\mathcal{G}_{111}) & V^{\prime}(\mathcal{G}_{1\bar{1}\bar{1}}+\mathcal{G}_{1\bar{1}3}) \\
	\\
	U\mathcal{G}_{11\bar{1}} & V(\mathcal{G}_{00\bar{2}} + \mathcal{G}_{220}) & V(\mathcal{G}_{20\bar{2}} + \mathcal{G}_{020}) & V(\mathcal{G}_{02\bar{2}}+\mathcal{G}_{200}) & V(\mathcal{G}_{000} + \mathcal{G}_{22\bar{2}}) & V^{\prime} (\mathcal{G}_{\bar{1}1\bar{1}}+\mathcal{G}_{31\bar{1}}) & V^{\prime}(\mathcal{G}_{1\bar{1}\bar{1}}+\mathcal{G}_{13\bar{1}}) & V^{\prime}(\mathcal{G}_{11\bar{3}}+\mathcal{G}_{111}) \\ \\
	
	U\mathcal{G}_{200} & V(\mathcal{G}_{1\bar{1}\bar{1}} + \mathcal{G}_{311}) & V(\mathcal{G}_{3\bar{1}\bar{1}} + \mathcal{G}_{111}) & V(\mathcal{G}_{11\bar{1}}+\mathcal{G}_{3\bar{1}1}) & V(\mathcal{G}_{1\bar{1}1} + \mathcal{G}_{31\bar{1}}) & V^{\prime} (\mathcal{G}_{000}+\mathcal{G}_{400}) & V^{\prime}(\mathcal{G}_{2\bar{2}0}+\mathcal{G}_{220}) & V^{\prime}(\mathcal{G}_{20\bar{2}}+\mathcal{G}_{202}) \\ \\
	
	U\mathcal{G}_{020} & V(\mathcal{G}_{\bar{1}1\bar{1}} + \mathcal{G}_{131}) & V(\mathcal{G}_{11\bar{1}} + \mathcal{G}_{\bar{1}31}) & V(\mathcal{G}_{\bar{1}3\bar{1}}+\mathcal{G}_{111}) & V(\mathcal{G}_{\bar{1}11} + \mathcal{G}_{13\bar{1}}) & V^{\prime} (\mathcal{G}_{\bar{2}20}+\mathcal{G}_{220}) & V^{\prime}(\mathcal{G}_{000}+\mathcal{G}_{040}) & V^{\prime}(\mathcal{G}_{02\bar{2}}+\mathcal{G}_{022}) \\ \\
	
	U\mathcal{G}_{002} & V(\mathcal{G}_{\bar{1}\bar{1}1} + \mathcal{G}_{113}) & V(\mathcal{G}_{1\bar{1}1} + \mathcal{G}_{\bar{1}13}) & V(\mathcal{G}_{\bar{1}11}+\mathcal{G}_{1\bar{1}3}) & V(\mathcal{G}_{\bar{1}\bar{1}3} + \mathcal{G}_{111}) & V^{\prime} (\mathcal{G}_{\bar{2}02}+\mathcal{G}_{202}) & V^{\prime}(\mathcal{G}_{0\bar{2}2}+\mathcal{G}_{022}) & V^{\prime}(\mathcal{G}_{000}+\mathcal{G}_{004})
	
    }\colvec[.75]{
    \tilde{\Phi}_{0}^{+}\\ \\ \tilde{\Phi}_{1}^{+}\\ \\ \tilde{\Phi}_{2}^{+}\\ \\ \tilde{\Phi}_{3}^{+}\\ \\ \tilde{\Phi}_{4}^{+} \\ \\ \tilde{\Phi}_{5}^{+} \\ \\ \tilde{\Phi}_{6}^{+} \\ \\ \tilde{\Phi}_{7}^{+}
    }
    =
    \colvec[.75]{
    \tilde{\Phi}_{0}^{+}\\ \\ \tilde{\Phi}_{1}^{+}\\ \\ \tilde{\Phi}_{2}^{+}\\ \\ \tilde{\Phi}_{3}^{+}\\ \\  \tilde{\Phi}_{4}^{+} \\ \\ \tilde{\Phi}_{5}^{+} \\ \\ \tilde{\Phi}_{6}^{+} \\ \\ \tilde{\Phi}_{7}^{+} 
    }
\end{equation}

\subsubsection*{Spin triplets}

\begin{equation}\label{appendix_triplet_sel_cons_eqn}
    \colvec[0.85]{
   
	V(\mathcal{G}_{000}-\mathcal{G}_{222}) & V(\mathcal{G}_{200} - \mathcal{G}_{022}) & V(\mathcal{G}_{020} - \mathcal{G}_{202}) & V(\mathcal{G}_{002} - \mathcal{G}_{220}) & V^{\prime} (\mathcal{G}_{\bar{1}11}-\mathcal{G}_{311}) & V^{\prime}(\mathcal{G}_{1\bar{1}1}-\mathcal{G}_{131}) & V^{\prime}(\mathcal{G}_{11\bar{1}}-\mathcal{G}_{113}) \\
	\\
	V(\mathcal{G}_{\bar{2}00} - \mathcal{G}_{022}) & V(\mathcal{G}_{000} - \mathcal{G}_{\bar{2}22}) & V(\mathcal{G}_{\bar{2}20}-\mathcal{G}_{002}) & V(\mathcal{G}_{\bar{2}02} - \mathcal{G}_{020}) & V^{\prime} (\mathcal{G}_{\bar{3}11}-\mathcal{G}_{111}) & V^{\prime}(\mathcal{G}_{\bar{1}\bar{1}1}-\mathcal{G}_{\bar{1}31}) & V^{\prime}(\mathcal{G}_{\bar{1}1\bar{1}}-\mathcal{G}_{\bar{1}13}) \\
	\\
	V(\mathcal{G}_{0\bar{2}0} - \mathcal{G}_{202}) & V(\mathcal{G}_{2\bar{2}0} - \mathcal{G}_{002}) & V(\mathcal{G}_{000} - \mathcal{G}_{2\bar{2}2}) & V(\mathcal{G}_{0\bar{2}2} - \mathcal{G}_{200}) & V^{\prime} (\mathcal{G}_{\bar{1}\bar{1}1}-\mathcal{G}_{3\bar{1}1}) & V^{\prime}(\mathcal{G}_{1\bar{3}1}-\mathcal{G}_{111}) & V^{\prime}(\mathcal{G}_{1\bar{1}\bar{1}}-\mathcal{G}_{1\bar{1}3}) \\
	\\
	V(\mathcal{G}_{00\bar{2}} - \mathcal{G}_{220}) & V(\mathcal{G}_{20\bar{2}} - \mathcal{G}_{020}) & V(\mathcal{G}_{02\bar{2}}-\mathcal{G}_{200}) & V(\mathcal{G}_{000} - \mathcal{G}_{22\bar{2}}) & V^{\prime} (\mathcal{G}_{\bar{1}1\bar{1}}-\mathcal{G}_{31\bar{1}}) & V^{\prime}(\mathcal{G}_{1\bar{1}\bar{1}}-\mathcal{G}_{13\bar{1}}) & V^{\prime}(\mathcal{G}_{11\bar{3}}-\mathcal{G}_{111}) \\ \\
	
	V(\mathcal{G}_{1\bar{1}\bar{1}} - \mathcal{G}_{311}) & V(\mathcal{G}_{3\bar{1}\bar{1}} - \mathcal{G}_{111}) & V(\mathcal{G}_{11\bar{1}}-\mathcal{G}_{3\bar{1}1}) & V(\mathcal{G}_{1\bar{1}1} - \mathcal{G}_{31\bar{1}}) & V^{\prime} (\mathcal{G}_{000}-\mathcal{G}_{400}) & V^{\prime}(\mathcal{G}_{2\bar{2}0}-\mathcal{G}_{220}) & V^{\prime}(\mathcal{G}_{20\bar{2}}-\mathcal{G}_{202}) \\ \\
	
	V(\mathcal{G}_{\bar{1}1\bar{1}} - \mathcal{G}_{131}) & V(\mathcal{G}_{11\bar{1}} - \mathcal{G}_{\bar{1}31}) & V(\mathcal{G}_{\bar{1}3\bar{1}}-\mathcal{G}_{111}) & V(\mathcal{G}_{\bar{1}11} - \mathcal{G}_{13\bar{1}}) & V^{\prime} (\mathcal{G}_{\bar{2}20}-\mathcal{G}_{220}) & V^{\prime}(\mathcal{G}_{000}-\mathcal{G}_{040}) & V^{\prime}(\mathcal{G}_{02\bar{2}}-\mathcal{G}_{022}) \\ \\
	
	V(\mathcal{G}_{\bar{1}\bar{1}1} - \mathcal{G}_{113}) & V(\mathcal{G}_{1\bar{1}1} - \mathcal{G}_{\bar{1}13}) & V(\mathcal{G}_{\bar{1}11}-\mathcal{G}_{1\bar{1}3}) & V(\mathcal{G}_{\bar{1}\bar{1}3} - \mathcal{G}_{111}) & V^{\prime} (\mathcal{G}_{\bar{2}02}-\mathcal{G}_{202}) & V^{\prime}(\mathcal{G}_{0\bar{2}2}-\mathcal{G}_{022}) & V^{\prime}(\mathcal{G}_{000}-\mathcal{G}_{004})
	
    }\colvec[.85]{
    \tilde{\Phi}_{1}^{-}\\ \\ \tilde{\Phi}_{2}^{-}\\ \\ \tilde{\Phi}_{3}^{-}\\ \\ \tilde{\Phi}_{4}^{-} \\ \\ \tilde{\Phi}_{5}^{-} \\ \\ \tilde{\Phi}_{6}^{-} \\ \\ \tilde{\Phi}_{7}^{-}
    }
    =
    \colvec[.85]{
    \tilde{\Phi}_{1}^{-}\\ \\ \tilde{\Phi}_{2}^{-}\\ \\ \tilde{\Phi}_{3}^{-}\\ \\ \tilde{\Phi}_{4}^{-} \\ \\ \tilde{\Phi}_{5}^{-} \\ \\ \tilde{\Phi}_{6}^{-} \\ \\ \tilde{\Phi}_{7}^{-} 
    }
\end{equation}
where 
\begin{equation}\label{greens_function_appendix}
    \mathcal{G}_{lmn}({\bf P}) = \frac{1}{N} \sum_{\bf q} \frac{e^{i(l\frac{q_{x}}{2}+m\frac{q_{y}}{2} + n\frac{q_{z}}{2} ) }}{E - \varepsilon_{\frac{\bf P}{2} + {\bf q}} -\varepsilon_{\frac{\bf P}{2} - {\bf q}} } = -\int_{-2\pi}^{2\pi}\int_{-2\pi}^{2\pi}\int_{-2\pi}^{2\pi} \frac{dq_{x} dq_{y} dq_{z}}{(4\pi)^{3}} \frac{\cos (l\frac{q_{x}}{2}+m\frac{q_{y}}{2}+n\frac{q_{z}}{2})}{|E| + \varepsilon_{\frac{\bf P}{2} + {\bf q} } + \varepsilon_{\frac{\bf P}{2} - {\bf q} }} \: .
\end{equation}

\section{Strong coupling}

In this section, we derive results for strong coupling cases for the chain, square and simple cubic lattices.

\subsection{Chain Lattice}

We start by considering the chain lattice with
\begin{equation}
{\bf b}_{1} = \left\{  a \right\} \, , \hspace{0.5cm} 
{\bf b}_{2} = \left\{ 2a \right\} \, .
\label{chain:dimer_vectors}
\end{equation}
Application of the Hamiltonian, $\hat{H}' = -t\sum_{\langle\nvec,{\bf b}_{1} \rangle \sigma} \hat{c}_{\nvec+{\bf b}_{1},\sigma}^{\dagger} \hat{c}_{\nvec\sigma} -t^{\prime} \sum_{\langle\nvec,{\bf b}_{2} \rangle\sigma} \hat{c}_{\nvec+{\bf b}_{2}, \sigma}^{\dagger} \hat{c}_{\nvec\sigma}$ in the space of strong coupling dimers (note that there is only hopping because the infinite $U$ avoids onsite pairs and strong attractive $V$ and $V'$ forces pairs onto NN and NNN sites) gives,
\begin{align}
    \hat{H}' D_{1,{\nvec}} & = - t^{\prime} \left( D_{1,{\nvec}-{\bf b}_1} + D_{1,{\nvec}+{\bf b}_1} \right) \\
    &\quad\quad - t \left( D_{2,{\nvec}} + D_{2,{\nvec}-{\bf b}_1} \right)\nonumber \\
    \hat{H}' D_{2,{\nvec}} & = -t \left( D_{1,{\nvec}} + D_{1,{\nvec}+{\bf b}_1} \right) \: .
\end{align}
In Fourier space this reduces to
\begin{equation}
    \colvec[.9]{
    E(P) + t^{\prime} \left(e^{iPa} + e^{-iPa} \right) & t\left(1 + e^{-iPa} \right)
    \\
    t\left(1 + e^{iPa} \right) & E(P) 
    }
    \colvec[1.1]{
    D_{1,P}\\D_{2,P}} = 0 \: ,
\end{equation}
which has solution
\begin{equation}
    E^{2} + 2t^{\prime}\cos{Pa}\cdot E - 4t^{2}\cos^{2}{\frac{Pa}{2}} = 0 \: ,
\end{equation}
with roots
\begin{equation}
    E(P) = -t^{\prime}\cos{Pa}\ \pm \left(t^{\prime 2}\cos^{2}{Pa} + 4t^{2}\cos^{2}{\frac{Pa}{2}} \right)^{\frac{1}{2}} .
    \label{eq:1d_sup_energy_arbtraryMomentum}
\end{equation}
This can be expanded in small $P$ to derive a generalized solution for the pair mass at any arbitrary NNN hopping $t^{\prime} = \theta t$, where $0 \le \theta \le 1$. The effective pair mass can be obtained directly using the expression
\begin{equation}\label{eq:str_coup_mass_arb_t_prime_1D}
    \left.\frac{m^{*}}{m_{0}}\right\vert_{1D} = \frac{2\left(1+4\theta\right)\sqrt{\theta^2 + 4 }}{\theta\left(\theta + \sqrt{\theta^2 + 4 } \right) + 1} \: .
\end{equation}

\subsection{Square Lattice}

For strongly bound NN and NNN intersite pairs on the square lattice, there are four dimers $D^{s}_{i,\nvec}$ between the sites
\begin{equation}
{\bf b}_{1} = \left\{  a, 0\right\} ,
{\bf b}_{2} = \left\{  a,-a \right\},
{\bf b}_{3} = \left\{  0,-a \right\},
{\bf b}_{4} = \left\{ -a,-a \right\} .
\label{square:dimer_vectors}
\end{equation}
Operating with the Hamiltonian on this strong coupling space leads to
\begin{align}
    \hat{H}' D_{1,{\nvec}} & = -t\left( D_{2,{\nvec}} + D_{2,{\nvec}-{\bf b}_{3}}\right) - t^{\prime} \left( D_{3,{\nvec}} + D_{3,{\nvec}+{\bf b}_1} + D_{3,{\nvec}-{\bf b}_3} + D_{3,{\nvec}-{\bf b}_4} \right) - t \left( D_{4,{\nvec}+{\bf b}_{1}} + D_{4,{\nvec}-{\bf b}_4} \right) \\
    \hat{H}' D_{2,{\nvec}} & = -t \left( D_{1,{\nvec}} + D_{1,{\nvec}+{\bf b}_3} \right) -t \left( D_{3,{\nvec}} + D_{3,{\nvec}+{\bf b}_1} \right) \\
    \hat{H}' D_{3,{\nvec}} & = -t^{\prime} \left( D_{1,{\nvec}} + D_{1,{\nvec}-{\bf b}_1} + D_{1,{\nvec}+{\bf b}_3} + D_{1,{\nvec}+{\bf b}_4} \right) -t \left( D_{2,{\nvec}} + D_{2,{\nvec}-{\bf b}_1} \right) -t \left(D_{4,{\nvec}} + D_{4,{\nvec}+{\bf b}_1} \right)  \\
    \hat{H}' D_{4,{\nvec}} & = -t \left( D_{1,{\nvec}-{\bf b}_{1}} + D_{1,{\nvec}+{\bf b}_4} \right) -t \left( D_{3,{\nvec}} + D_{3,{\nvec}-{\bf b}_1} \right) \: ,
\end{align}
which in Fourier space is
\begin{equation}
    \colvec[.92]{
    E({\bf P}) & t\left( 1 + e^{iP_{y}a} \right) & t^{\prime} \left(1 + e^{iP_{x}a} \right) \left( 1 + e^{iP_{y}a} \right) & t\left(1 + e^{iP_{y}a} \right)e^{iP_{x}a}
    \\
    t\left(1 + e^{-iP_{y}a} \right) & E({\bf P}) & t\left(1 + e^{iP_{x}a} \right) & 0 \\
    t^{\prime}\left(1 + e^{-iP_{x}a} \right)\left(1 + e^{-iP_{y}a} \right) & t\left(1 + e^{-iP_{x}a} \right) & E({\bf P}) & t\left(1 + e^{iP_{x}a} \right) \\
    t\left(1 + e^{-iP_{y}a} \right)e^{-iP_{x}a} & 0 & t\left(1 + e^{-iP_{x}a} \right) & E({\bf P})
    }
    \colvec[1.1]{
    D_{1,{\bf P}} \\
    D_{2,{\bf P}} \\
    D_{3,{\bf P}} \\
    D_{4,{\bf P}} } = 0 \: .
\end{equation}
For small pair momentum ${\bf P} = (P_{x}, 0)$, the determinant yields
\begin{equation}
    E^{4} - \left[4t^{2}\left(3 + \cos{P_{x}a} \right) + 8t^{\prime2}\left(1 + \cos{P_{x}a} \right) \right] \cdot E^{2} + 32t^{2}t^{\prime}\left[ 1 + \cos{P_{x}a} \right] \cdot E  = 0 \: .
    \label{eqn:scstrongdispersion}
\end{equation}
To derive the effective mass with $t^{\prime} = \theta t$, where $0 \le \theta \le 1$, we substitute $E = E_0 + \Delta$, where $\Delta = \frac{\hbar^2 P^2_x}{2m^{\ast}_{x}} \ll E_0$, and expand Eq.~(\ref{eqn:scstrongdispersion}) to the first order in $P^2_x$ and $\Delta$. After some algebra, we obtain
\begin{equation}\label{eq:str_coup_mass_arb_t_prime_2D}
    \left.\frac{m^{*}_{x}}{m_{0}}\right\vert_{SQ} = \frac{2(1+2\theta)(\theta^{2}+3\theta\sqrt{\theta^{2}+4} + 4)}{2\theta^{3}+5\theta + \sqrt{\theta^{2} + 4}(2\theta^2 + 1)} \: .
\end{equation}

\subsection{BCC Lattice}

For strong intersite pairs on the BCC lattice, there are dimers between the sites
\begin{align*}
{\bf b}_{1} = & \left\{ \frac{a}{2} ,   \frac{a}{2} ,   \frac{a}{2} \right\},
{\bf b}_{2} =   \left\{ \frac{a}{2} ,   \frac{a}{2} , - \frac{a}{2} \right\},
{\bf b}_{3} =   \left\{ \frac{a}{2} , - \frac{a}{2} ,   \frac{a}{2} \right\},\\ 
{\bf b}_{4} = & \left\{ \frac{a}{2} , - \frac{a}{2} , - \frac{a}{2} \right\},  
{\bf b}_{5} =   \left\{ a , 0 , 0 \right\},  
{\bf b}_{6} =   \left\{ 0 , a , 0 \right\},  \\  
{\bf b}_{7} = & \left\{ 0 , 0 , a \right\}  \, . 
\label{bcc:dimer_vectors}
\end{align*}
After applying the Hamiltonian operator in the dimer lattice subspace and Fourier transforming, we obtain
\begin{equation}
    \colvec[.75]{
    E({\bf P})  & t^{\prime}(1 + e^{i{\bf P}{\bf b}_7}) & t^{\prime} ( 1 + e^{i{\bf P}{\bf b}_6} )   &  t^{\prime} ( e^{i{\bf P}{\bf b}_1} + e^{-i{\bf P}{\bf b}_4} )  
            &   t ( 1 + e^{-i{\bf P}{\bf b}_4} )   &  t ( 1 + e^{i{\bf P}{\bf b}_3} ) &  t ( 1 + e^{i{\bf P}{\bf b}_2} ) \\
    t^{\prime}(1 + e^{-i{\bf P}{\bf b}_7}) & E({\bf P}) & t^{\prime} ( e^{i{\bf P}{\bf b}_2} + e^{-i{\bf P}{\bf b}_3} )  &  t^{\prime} ( 1 + e^{i{\bf P}{\bf b}_6} )  
            &   t ( 1 + e^{-i{\bf P}{\bf b}_3} )  
            &   t ( 1 + e^{ i{\bf P}{\bf b}_4} )  
            &   t (e^{i{\bf P}{\bf b}_2} + e^{-i{\bf P}{\bf b}_7} )    \\
    t^{\prime} ( 1 + e^{-i{\bf P}{\bf b}_6} )  &   t^{\prime} (  e^{-i{\bf P}{\bf b}_2} + e^{i{\bf P}{\bf b}_3} )  &  E({\bf P})  &   t^{\prime} ( 1 + e^{ i{\bf P}{\bf b}_7} )
            &   t ( 1 + e^{-i{\bf P}{\bf b}_2} )   
            &   t ( e^{i{\bf P}{\bf b}_3} + e^{-i{\bf P}{\bf b}_6} )      
            &   t ( 1 + e^{ i{\bf P}{\bf b}_4} ) \\            
t^{\prime} ( e^{-i{\bf P}{\bf b}_1} + e^{i{\bf P}{\bf b}_4} )  &   t^{\prime} ( 1 + e^{-i{\bf P}{\bf b}_6} )  &  t^{\prime}(1 + e^{-i{\bf P}{\bf b}_7} )  &  E({\bf P}) 
            &   t ( 1 + e^{-i{\bf P}{\bf b}_1} )    &   t ( e^{i{\bf P}{\bf b}_4} + e^{-i{\bf P}{\bf b}_6} )   &   t ( e^{i{\bf P}{\bf b}_4} + e^{-i{\bf P}{\bf b}_7} )   \\
t ( 1 + e^{i{\bf P}{\bf b}_4} )  &  t ( 1 + e^{i{\bf P}{\bf b}_3} )
            &  t ( 1 + e^{i{\bf P}{\bf b}_2} )  
            &  t ( 1 + e^{i{\bf P}{\bf b}_1} )   &  E({\bf P}) &  0  & 0   \\
t ( 1 + e^{-i{\bf P}{\bf b}_3} )  &  t ( 1 + e^{-i{\bf P}{\bf b}_4} ) 
            &  t ( e^{i{\bf P}{\bf b}_6} + e^{-i{\bf P}{\bf b}_3} )
            &  t ( e^{i{\bf P}{\bf b}_6} + e^{-i{\bf P}{\bf b}_4} )    &   0   &   E({\bf P})  & 0 \\
t ( 1 + e^{-i{\bf P}{\bf b}_2} )  &  t ( e^{i{\bf P}{\bf b}_7} + e^{-i{\bf P}{\bf b}_2} ) 
            &  t ( 1 + e^{-i{\bf P}{\bf b}_4} )
            &  t (     e^{ i{\bf P}{\bf b}_7} + e^{-i{\bf P}{\bf b}_4} )    &   0  & 0  &   E({\bf P})        
    }
    \colvec[.9]{D_{1,{\bf P}} \\
                D_{2,{\bf P}} \\
                D_{3,{\bf P}} \\
                D_{4,{\bf P}} \\
                D_{5,{\bf P}} \\
                D_{6,{\bf P}} \\
                D_{7,{\bf P}}
    } = 0 \: . \label{bcctwopart:eq:appgtwelve}
\end{equation}
For ${\bf P} = (P_{x},0,0)$, the expansion of the determinant in Eq.~(\ref{bcctwopart:eq:appgtwelve}) gives
\begin{align}
\begin{split}
    & E^{7} - \left[24t^{2}\left(1 + \cos{\frac{P_{x}a}{2}}\right) + 4t^{\prime\;2}\left(5 + \cos{P_{x}a} \right)\right]E^{5} + \left[24t^{2}t^{\prime}\left(5 + \cos{P_{x}a} + 6\cos{\frac{P_{x}a}{2}} \right) + 64t^{\prime\;3}\cos{\frac{P_{x}a}{2}} \right]E^{4} \\
    & + \left[24t^{2}t^{\prime\;2}\left(\cos{\frac{3P_{x}a}{2}} -6\cos{P_{x}a} - 13\cos{\frac{P_{x}a}{2}} - 6 \right) + 2t^{\prime\;4}\left(\cos{2P_{x}a} - 12\cos{P_{x}a} - 13 \right) \right]E^{3} \\ 
    & + 24 t^{2} t^{\prime\;3} \left[-\cos{2P_{x}a} + 2\cos{\frac{3P_{x}a}{2}} + 4\cos{P_{x}a} + 8\cos{\frac{P_{x}a}{2}} + 5 \right] E^{2} = 0 \: .
    \end{split}
\end{align}
We note flat bands with $E=0$. After dividing by $E^2$, we obtain a quintic that cannot be solved analytically. 

We can find analytic values for the special case $t^{\prime} = 0$. In that case, there are only two non-zero solutions of $E$ which are given by the equation
\begin{equation}
    E(P_{x}) = \pm \sqrt{24t^{2}\left( 1 + \cos{\frac{P_{x}a}{2}} \right)} \: ,
\end{equation}
where the lowest state at small pair momenta has the energy 
\begin{equation}
    E(P_{x}a \ll 1) \approx - 4 \sqrt{3} \, t + \frac{\sqrt{3}}{8}(P_{x}a)^{2}t \equiv E_{0} + \frac{\hbar^{2}P_{x}^{2}}{2m^{*}_{x}}
\end{equation}
The pair mass, expressed in terms of the bare mass, is thus
\begin{equation}\label{eqapp:sup_mass_t_prime0_BCC}
    m^{*}_x = \frac{4\hbar^{2}}{a^{2}t\sqrt{3}} = \frac{8}{\sqrt{3}}m_{0}= 4.618802\dots m_{0} \: .
\end{equation}



\end{widetext}

\bibliographystyle{unsrt}
\bibliography{References}

\begin{thebibliography}{10}

\bibitem{Uemura1989}
Y.J. Uemura, G.M. Luke, B.J. Sternlieb, J.H. Brewer, J.F. Carolan, W.N. Hardy,
  R.~Kadono, J.R. Kempton, R.F. Kiefl, S.R. Kreitzman, P.~Mulhern, T.M.
  Riseman, D.Ll. Williams, B.X. Yang, S.~Uchida, H.~Takagi, J.~Gopalakrishnan,
  A.W. Sleight, M.A. Subramanian, C.L. Chien, M.Z. Cieplak, Gang Xiao, V.Y.
  Lee, B.W. Statt, C.E. Stronach, W.J. Kossler, and X.H. Yu.
\newblock Universal correlations between {$T_c$} and $n_{s}/m^{\ast}$ (carrier
  density over effective mass) in high-{$T_c$} cuprate superconductors.
\newblock {\em Physical Review Letters}, 62:2317--2320, 1989.

\bibitem{Uemura1991}
Y.J. Uemura, L.P. Le, G.M. Luke, B.J. Sternlieb, W.D. Wu, J.H. Brewer, T.M.
  Riseman, C.L. Seaman, M.B. Maple, M.~Ishikawa, D.G. Hinks, J.D. Jorgensen,
  G.~Saito, and H.~Yamochi.
\newblock Basic similarities among cuprate, bismuthate, organic,
  {C}hevrel-phase, and heavy-fermion superconductors shown by penetration-depth
  measurements.
\newblock {\em Physical Review Letters}, 66:2665--2668, 1991.

\bibitem{Alexandrov1994}
A.S. Alexandrov and N.F. Mott.
\newblock {\em High Temperature Superconductors and other Superfluids}.
\newblock Taylor \& Francis, London, 1994.

\bibitem{Alexandrov1999b}
A.S. Alexandrov and V.V. Kabanov.
\newblock {Parameter-free expression for superconducting $T_{c}$ in cuprates}.
\newblock {\em Phys. Rev. B}, 59:13628--13631, 1999.

\bibitem{Micnas1990}
R.~Micnas, J.~Ranninger, and S.~Robaszkiewicz.
\newblock Superconductivity in narrow-band systems with local nonretarded
  attractive interactions.
\newblock {\em Reviews of Modern Phy\-sics}, 62:113--171, 1990.

\bibitem{Seo2019}
Y.I. Seo, W.J. Choi, S.-I. Kimura, and Y.S. Kwon.
\newblock Evidence for a preformed cooper pair model in the pseudogap spectra
  of a {Ca$_{10}$(Pt$_4$As$_8$)(Fe$_2$As$_2$)$_5$} single crystal with a nodal
  superconducting gap.
\newblock {\em Scientific Reports}, 9:3987, 2019.

\bibitem{Kang2020}
B.L. Kang, M.Z. Shi, S.J. Li, H.H. Wang, Q.~Zhang, D.~Zhao, J.~Li, D.W. Song,
  L.X. Zheng, L.P. Nie, T.~Wu, and X.H. Chen.
\newblock Preformed {C}ooper pairs in layered {FeSe}-based superconductors.
\newblock {\em Phys. Rev. Lett.}, 125:097003, 2020.

\bibitem{Zhou2019}
Panpan Zhou, Liyang Chen, Yue Liu, Ilya Sochnikov, Anthony~T. Bollinger,
  Myung-Geun Han, Yimei Zhu, Xi~He, Ivan Boz\'ovi\u{c}, and Douglas Natelson.
\newblock Electron pairing in the pseudogap state revealed by shot noise in
  copper oxide junctions.
\newblock {\em Nature}, 572:493--496, 2019.

\bibitem{Ivanov1994}
V.A. Ivanov, P.E. Kornilovitch, and V.V. Bobryshev.
\newblock Possible enhancement of the phonon pairing mechanism by a staggered
  magnetic field.
\newblock {\em Physica C}, 235-240:2369--2370, 1994.

\bibitem{Kornilovitch2015}
P.E. Kornilovitch and J.P. Hague.
\newblock Optimal interlayer hopping and high temperature {Bose--Einstein}
  condensation of local pairs in quasi {2D} superconductors.
\newblock {\em J. Phys.: Condens. Matter}, 27:075602, 2015.

\bibitem{Zhang2022}
C.~Zhang, N.V. Prokof'ev, and B.V. Svistunov.
\newblock Bond bipolarons: Sign-free {Monte Carlo} approach.
\newblock {\em Phys. Rev. B}, 105:L020501, 2022.

\bibitem{Emery1990}
V.J. Emery, S.A. Kivelson, and H.Q. Lin.
\newblock Phase separation in the {$t-{J}$} model.
\newblock {\em Phys. Rev. Lett.}, 64:475--478, 1990.

\bibitem{Dagotto1993}
E.~Dagotto and J.~Riera.
\newblock Indications of $d_{x^2-d^2}$ superconductivity in the two dimensional
  {$t-J$} model.
\newblock {\em Phys. Rev. Lett.}, 70:682, 1993.

\bibitem{Kornilovitch2013}
P.E. Kornilovitch.
\newblock Stability of three-fermion clusters with finite range of attraction.
\newblock {\em Europhys. Lett.}, 103:27005, 2013.

\bibitem{Kornilovitch2014}
P.E. Kornilovitch.
\newblock Ferromagnetism and {B}orromean binding in three-fermion clusters.
\newblock {\em Phys. Rev. Lett.}, 112:077202, 2014.

\bibitem{Chakraborty2014}
M.~Chakraborty, M.~Tezuka, and B.I. Min.
\newblock Interacting-{H}olstein and extended-{H}olstein bipolarons.
\newblock {\em Phys. Rev. B}, 89:035106, 2014.

\bibitem{Kornilovitch2020}
P.E. Kornilovitch.
\newblock Trion formation and unconventional superconductivity in a
  three-dimensional model with short-range attraction.
\newblock {\em Int. J. Mod. Phys. B}, 23:2050042, 2020.

\bibitem{Kornilovitch2022}
P.~Kornilovitch.
\newblock Stable pair liquid phase in fermionic systems.
\newblock {\em Phys. Rev. B}, 107:115135, 2023.

\bibitem{hague2007superlighta}
J.P. Hague, P.E. Kornilovitch, J.H. Samson, and A.S. Alexandrov.
\newblock Superlight small bipolarons.
\newblock {\em J. Phys.: Condens. Matter}, 19:255214, 2007.

\bibitem{hague2008_sing_trip_bip_triangular}
J.P. Hague, P.E. Kornilovitch, J.H. Samson, and A.S. Alexandrov.
\newblock Singlet and triplet bipolarons on the triangular lattice.
\newblock {\em J. Phys. Chem. Solids}, 69:3304--3306, 2008.

\bibitem{alexandrov2002b}
A~S Alexandrov and P~E Kornilovitch.
\newblock The {F}r\"{o}hlich-{C}oulomb model of high-temperature
  superconductivity and charge segregation in the cuprates.
\newblock {\em J. Phys.: Condens. Matter}, 14:5337, 2002.

\bibitem{Kornilovitch2023}
P.E. Kornilovitch.
\newblock Two-particle bound states on a lattice.
\newblock {\em Annals of Physics}, 460:169574, 2024.

\bibitem{adebanjofcc2022}
G.D. Adebanjo, P.E. Kornilovitch, and J.P. Hague.
\newblock Superlight pairs in face-centred-cubic extended {H}ubbard models with
  strong {C}oulomb repulsion.
\newblock {\em J. Phys.: Condens. Matter}, 34:135601, 2022.

\bibitem{Callaway1989}
J.~Callaway J, D.G. Kanhere, and P.K. Misra.
\newblock Polarization-induced pairing in high-temperature superconductivity.
\newblock {\em Physical Review B}, 36:7141--7144, 1987.

\bibitem{Bussmann1989}
A.~Bussmann-Holder, A.~Simon, and H.~B\"uttner.
\newblock Possibility of a common origin to ferroelectricity and
  superconductivity in oxides.
\newblock {\em Physical Review B}, 39:207--214, 1989.

\bibitem{Zhang1991}
X.~Zhang and C.R.A. Catlow.
\newblock Elastic and {C}oulombic contributions to real-space hole pairing in
  doped {La$_2$CuO$_4$}.
\newblock {\em J. Mater. Chem.}, 1:233, 1991.

\bibitem{Catlow1998}
C.R.A. Catlow, M.S. Islam, and X.~Zhang.
\newblock The structure and energies of peroxy bipolarons in
  {L}a$_2${C}u{O}$_4$.
\newblock {\em J. Phys.: Condens. Matter}, 10:L49, 1998.

\bibitem{Mihailovich2001}
D.~Mihailovic and V.V. Kabanov.
\newblock {Finite wave vector Jahn-Teller pairing and superconductivity in the
  cuprates}.
\newblock {\em Physical Review B}, 63:054505, 2001.

\bibitem{Kabanov2002}
V.V. Kabanov and D.~Mihailovic.
\newblock Manifestations of mesoscopic {J}ahn-{T}eller real-space pairing and
  clustering in {YBa$_2$Cu$_3$O$_{7-\delta}$}.
\newblock {\em Physical Review B}, 65:212508, 2002.

\bibitem{Scalapino2012}
D.J. Scalapino.
\newblock A common thread: The pairing interaction for unconventional
  superconductors.
\newblock {\em Rev. Mod. Phys.}, 84:1383--1417, 2012.

\bibitem{alexandrov2013SCTHTSC}
A.~S. Alexandrov.
\newblock {\em Strong-Coupling Theory of High-Temperature Superconductivity}.
\newblock Cambridge University Press, 2013.

\bibitem{hague2007staggeredladder}
J.P. Hague, P.E. Kornilovitch, J.H. Samson, and A.S. Alexandrov.
\newblock Long-range electron-phonon interactions lead to superlight small
  bipolarons.
\newblock {\em J. Phys.: Conf. Ser.}, 92:012118, 2007.

\bibitem{Dagotto1994}
E.~Dagotto.
\newblock Correlated electrons in high-temperature superconductors.
\newblock {\em Rev. Mod. Phys.}, 66:763--840, 1994.

\bibitem{Leung1995}
P.W. Leung and R.J. Gooding.
\newblock Dynamical properties of the single-hole {$t-J$} model on a 32-site
  square cluster.
\newblock {\em Phys. Rev. B}, 52:R15711--R15714, 1995.

\bibitem{Note1}
A dilute Hamiltonian is easily determined, since at very large $U$ and
  half-filling the wavefunction is a Fock state with 1 particle per site (with
  constant energy determined by the intersite interactions). When holes are
  introduced to this Fock state, the difference in the Hamiltonian from the
  constant background is equivalent to the dilute Hamiltonian. For smaller $U$,
  such an approximation becomes less good as double occupancy is permitted and
  increased kinetic energy contributions complicate the half-filled
  wavefunction.

\bibitem{hardy2009}
T.~M. Hardy, J.~P. Hague, J.~H. Samson, and A.~S. Alexandrov.
\newblock Superconductivity in a hubbard-fr\"ohlich model and in cuprates.
\newblock {\em Phys. Rev. B}, 79:212501, Jun 2009.

\bibitem{chen2021a}
Z.~Chen, Y.~Wang, S.~N. Rebec, T.~Jia, M.~Hashimoto, D.~Lu, B.~Moritz, R.~G.
  Moore, T.~P. Devereaux, and Z-X Shen.
\newblock Anomalously strong near-neighbor attraction in doped {1D} cuprate
  chains.
\newblock {\em Science}, 373:1235--1239, 2021.

\bibitem{wang_chen_et_al_2021}
Y.~Wang, Z.~Chen, T.~Shi, B.~Moritz, Z-X Shen, and T.~P. Devereaux.
\newblock Phonon-mediated long-range attractive interaction in one-dimensional
  cuprates.
\newblock {\em Phys. Rev. Lett.}, 127:197003, 2021.

\bibitem{jozef2017}
J.~Spa\l{}ek, M.~Zegrodnik, and J.~Kaczmarczyk.
\newblock Universal properties of high-temperature superconductors from
  real-space pairing: {$t$-$J$-$U$} model and its quantitative comparison with
  experiment.
\newblock {\em Phys. Rev. B}, 95:024506, 2017.

\bibitem{hubbard1963electron}
J.~Hubbard.
\newblock Electron correlations in narrow energy bands.
\newblock {\em Proc. R. Soc. Lond. A}, 276:238--257, 1963.

\bibitem{kato2000}
T.~Kato and M.~Kato.
\newblock Stripe orders in the extended {Hubbard} model.
\newblock {\em J. Phys. Soc. Japan}, 69:3972--3979, 2000.

\bibitem{PhysRevB.97.184507}
M.~Jiang, U.~R. H\"ahner, T.~C. Schulthess, and T.~A. Maier.
\newblock $d$-wave superconductivity in the presence of nearest-neighbor
  {Coulomb} repulsion.
\newblock {\em Phys. Rev. B}, 97:184507, 2018.

\bibitem{Laad_1991}
M.~S. Laad and D.~K. Ghosh.
\newblock Extended {Hubbard} model in two dimensions.
\newblock {\em J. Phys.: Condens. Matter}, 3:9723--9732, 1991.

\bibitem{Micnas_1988}
R.~Micnas, J.~Ranninger, and S.~Robaszkiewicz.
\newblock An extended {H}ubbard model with inter-site attraction in two
  dimensions and high-{$T_c$} superconductivity.
\newblock {\em J. Phys. C: Solid State Physics}, 21:L145--L151, 1988.

\bibitem{Zhang1988}
F.C. Zhang and T.M. Rice.
\newblock Effective {H}amiltonian for the superconducting {Cu} oxides.
\newblock {\em Physical Review B}, 37:3759, 1988.

\bibitem{Sakakibara2010}
H.~Sakakibara, H.~Usui, K.~Kuroki, R.~Arita, and H.~Aoki.
\newblock Two-orbital model explains the higher transition temperature of the
  single-layer {Hg}-cuprate superconductor compared to that of the {La}-cuprate
  superconductor.
\newblock {\em Physical Review Letters}, 105:057003, 2010.

\bibitem{Hirayama2018}
M.~Hirayama, Y.~Yamaji, T.~Misawa, and M.~Imada.
\newblock {\em Ab initio} effective hamiltonians for cuprate superconductors.
\newblock {\em Physical Review B}, 98:134501, 2018.

\bibitem{Jiang2023}
Sh. Jiang, D.J. Scalapino, and S.R. White.
\newblock Density matrix renormalization group based downfolding of the
  three-band hubbard model: Importance of density-assisted hopping.
\newblock {\em Physical Review B}, 108:L161111, 2023.

\bibitem{Lin1991}
H.Q. Lin.
\newblock Dilute gas of electron pairs in the {$t-J$} model.
\newblock {\em Physical Review B}, 44:4674--4676, 1991.

\bibitem{Petukhov1992}
A.G. Petukhov, J.~Gal\'an, and J.A. Verg\'es.
\newblock Bound states of two electrons described by the {$t-J$} model.
\newblock {\em Physical Review B}, 46:6212--6216, 1992.

\bibitem{Kagan1994}
M.Yu. Kagan and T.M. Rice.
\newblock Superconductivity in the two-dimensional {$t-J$} model at low
  electron density.
\newblock {\em J. Phys.: Condens. Matter}, 6:3771--3780, 1994.

\bibitem{Chernyshev1999}
A.L. Chernyshev and P.W. Leung.
\newblock Holes in the {$t-J_{z}$} model: A diagrammatic study.
\newblock {\em Physical Review B}, 60:1592--1606, 1999.

\bibitem{kornilovitch2004}
P.E. Kornilovitch.
\newblock Enhanced stability of bound pairs at nonzero lattice momenta.
\newblock {\em Phys. Rev. B}, 69:235110, 2004.

\bibitem{martin1993}
R.L. Martin and J.P. Ritchie.
\newblock Coulomb and exchange interactions in {C}$^{n-}_{60}$.
\newblock {\em Phys. Rev. B}, 48:4845, 1993.

\bibitem{Bak1999}
M.~Bak and R.~Micnas.
\newblock Extended bound states of fermions on 2d square lattice beyond the nn
  hopping and interactions.
\newblock {\em Molecular Physics Reports}, 24:168--176, 1999.

\bibitem{harrison2024}
N.~Harrison and M.K. Chan.
\newblock Thermodynamic evidence for electron correlation-driven flattening of
  the quasiparticle bands in the high-$t_{C}$ cuprates.
\newblock arXiv:2303.12956.

\bibitem{Leggett2006}
A.~J. Leggett.
\newblock What do we know about high ${T}_{c}$?
\newblock {\em Nature Physics}, 2:134--136, 2006.

\bibitem{bogoliubov}
N.N. Bogoliubov.
\newblock {\em Lectures on Quantum Statistics, Volume 2 Quasi-averages}.
\newblock Gordon and Breach, Science Publishers Ltd., London and New York,
  1970.

\bibitem{hague2006dwave}
J.~P. Hague.
\newblock $d$-wave superconductivity from electron-phonon interactions.
\newblock {\em Phys. Rev. B}, 73:060503(R), Feb 2006.

\bibitem{gunnarsson2004alkali}
O.~Gunnarsson.
\newblock {\em Alkali-doped fullerides: narrow-band solids with unusual
  properties}.
\newblock World Scientific, Singapore, 2004.

\end{thebibliography}

\end{document}